\documentclass[a4paper,11pt]{article}

\usepackage[T1]{fontenc}
\usepackage[english]{babel}
\usepackage{enumerate}
\usepackage{hyperref}
\hypersetup{
	colorlinks=true,
	citecolor=blue,
	linkcolor=blue,
	filecolor=red,
	urlcolor=blue,
}
\usepackage{doi}
\usepackage{ragged2e}
\usepackage{setspace}
\usepackage{graphics}
\usepackage{paralist}
\usepackage[numbers,sort&compress]{natbib}
\usepackage{mathrsfs}
\usepackage{amsmath,amssymb,amsfonts}
\usepackage{graphicx}
\usepackage{epsfig}
\usepackage{epstopdf}
\usepackage[usenames,dvipsnames]{color}
\usepackage{subfigure}
\usepackage{appendix}
\title{Energy Extraction and Particle Acceleration in String-Inspired Rotating Einstein-Maxwell-Dilaton-Axion Black Hole}
\author{
	{Arindam Kumar Chatterjee$^{a}$\thanks{72arindam@gmail.com}},
	\\[0.3cm]
	{\small $^a$ Department of Physics, Gurukul Kangri (Deemed to be University), Haridwar - 249407, India.}\\[0.1cm]
}
\begin{document}
	\baselineskip=0.5 cm
	\date{}
	\maketitle
	
	\begin{abstract}
		\baselineskip=0.5 cm
		\noindent
		We study energy extraction and particle acceleration in the rotating Einstein-Maxwell-Dilaton-Axion (EMDA) black hole, focusing on the impact of dilaton hair $b\le 0$ on near-horizon energetics relative to Kerr. For the Penrose process we derive analytic expressions for the maximum efficiency and show that negative $b$ can strongly enhance the ideal gain in the extremal regime (e.g., reaching $\sim 91\%$ for $b=-0.3$). We then compute the irreducible mass $M_{\rm irr}$ and the corresponding rotationally extractable energy $\mathcal{E}_{\rm rot}\equiv M-M_{\rm irr}$, finding that $M_{\rm irr}$ decreases monotonically as $b$ becomes more negative while $\mathcal{E}_{\rm rot}$ increases, indicating a larger spin-energy reservoir; at extremality the extracted share from rotation is $\mathcal{E}_{\rm rot}/M\simeq 0.63$ for EMDA, reducing to the Kerr value $\simeq 0.29$ at $b=0$. Kinematic constraints relevant to fragment production are quantified via the Wald and Bardeen--Press--Teukolsky bounds, which are progressively relaxed for more negative $b$. For wave superradiance we obtain the flux balance and the amplification window $0<\beta<k\Omega_H$, with $\Omega_H$ expressed through $\Xi=r_H^{2}+2br_H+a^{2}$; negative $b$ modifies $\Omega_H$ and enlarges the parameter region exhibiting negative horizon flux. Finally, we analyse two-particle collisions and derive $E_{\rm cm}$, showing that the Ba\~nados--Silk--West divergence persists at the horizon when one particle is tuned to the critical angular momentum $L_c=E/\Omega_H$, while $E_{\rm cm}$ remains finite for generic angular momenta. Overall, dilaton hair in EMDA simultaneously amplifies energy-extraction channels and reshapes the near-horizon thresholds governing high-energy collisions.
	\end{abstract}
	
\section{Introduction}
In the last decades, cosmological observations indicate that nearly 95\% of the Universe consists of dark matter and dark energy, whose fundamental nature remains unresolved. Among the theoretical frameworks proposed to address these questions, string theory offers a promising route towards a consistent description of quantum gravity. In its low-energy regime, the massless sector of heterotic string theory can be described by Einstein--Maxwell--dilaton--axion (EMDA) gravity, which extends Einstein--Maxwell theory through the inclusion of the dilaton and axion scalar fields. These additional degrees of freedom may also have implications for dark-sector phenomenology~\citep{2011ASSL..370.....M}. Astrophysical observations further suggest that most galaxies harbour supermassive black holes at their centers. Since these objects acquire angular momentum through galactic dynamics and accretion, they are generally expected to be rotating. The determination of their mass, spin, and surrounding geometry is therefore of considerable theoretical and observational interest, particularly in the case of Sagittarius~A* (Sgr A*) at the Galactic Center. Dark matter, which constitutes approximately 27\% of the total cosmic energy density, is supported indirectly by galactic rotation curves~\citep{1980ApJ...238..471R}, galaxy-cluster dynamics~\citep{2009GReGr..41..207Z}, and cosmic microwave background anisotropies measured by the Planck mission~\citep{2014A&A...571A..16P}. At the same time, the singularity theorems of classical general relativity imply that spacetime singularities arise under broad physical conditions~\citep{PhysRevLett.14.57,Senovilla:2014gza,Landsman:2022hrn}. Such singularities are generally regarded as indications of the breakdown of the classical theory rather than as physically admissible structures. In the absence of a complete theory of quantum gravity, considerable attention has therefore been devoted to phenomenological extensions of general relativity and modified black hole geometries. The study of classical black holes, string-inspired solutions, and nonsingular alternatives consequently provides an important framework for examining strong-field gravity, black hole interiors, and possible departures from the standard Kerr paradigm.
	
\vspace{0.2cm}
\noindent
The rotating Einstein--Maxwell--dilaton--axion (EMDA) black hole arises as a solution of the low-energy effective theory of heterotic string theory~\citep{PhysRevLett.74.1276}. In contrast to the Kerr geometry, which is a vacuum solution of Einstein's field equations characterised only by mass and angular momentum, the EMDA spacetime incorporates an electromagnetic field together with the dilaton and axion fields. These additional degrees of freedom introduce a dilaton-charge parameter that modifies the metric, the electromagnetic coupling, and the causal structure of the spacetime, including the locations of the event horizon and ergoregion. The axion field further affects the electromagnetic duality structure and contributes additional interactions absent in the Kerr solution. The Kerr geometry is recovered in the appropriate limit in which the additional charges and scalar fields vanish. The EMDA black hole therefore provides a useful framework for investigating string-inspired modifications to black hole geometry, dynamics, and observable phenomena. Several properties of the rotating EMDA spacetime have been investigated in the literature. Jing~\citep{Jing:1996np} analysed its thermodynamic behaviour, including aspects of stability and phase structure. Pan~\citep{Pan:2007gs} numerically determined the fundamental and higher-overtone quasinormal frequencies of massless scalar perturbations using Leaver's continued-fraction method. The shadow of the EMDA black hole was studied by Wei~\citep{Wei_2013}, who showed that the dilaton and axion fields produce observable deviations from the Kerr shadow. Flathmann~\citep{Flathmann_2015} obtained exact analytical solutions of the separated geodesic equations in terms of elliptic functions and classified the corresponding bound, escape, and plunging trajectories. Collectively, these studies demonstrate that the dilaton and axion fields can significantly modify the thermodynamic, perturbative, optical, and orbital properties of the black hole spacetime.

\vspace{0.2cm}
\noindent
A seminal classical mechanism for extracting rotational energy from a black hole is the Penrose process~\citep{Penrose1971uk,chandrasekhar1983mathematical}. In this process, a particle entering the ergoregion splits or interacts such that one fragment occupies a negative-energy orbit, as measured by an observer at infinity, while the other escapes with an energy exceeding that of the incident particle. Such negative-energy states arise from frame dragging within the ergoregion and demonstrate that the rotational energy of a black hole can, in principle, be extracted. The Penrose process has subsequently been investigated in a wide variety of black hole spacetimes~\citep{chen2025electricpenroseprocessayonbeatogarcia,fatima2025revisitingpenroseprocessrotating,Abbasi:2025,Pradhan_2019,PhysRevD.104.084059,PhysRevD.110.123035,Mukherjee_2019,PhysRevD.89.024023,1985JApA,Rudra:2019ssz,1986ApJ,ganguly2014,PhysRevD8404,Abdujabbarov_2011,Chen:2013vja,Ortiqboev:2026mip,Ashraf:2025lxs,Ortiqboev:2025jzu,Ortiqboev:2024mtk}. A complementary wave-based mechanism is black hole superradiance, in which suitable field modes are amplified upon scattering from a rotating horizon, causing the reflected wave to carry greater energy than the incident wave~\citep{pengpan2025superradiancequasinormalmodesmassive,Mollicone_2025,Liu_2024,Jha_2024,Mukherjee_2019,Laurent_Di_Menza_2015,Cheriyodathillathu_2025,herrerovalea2024superradiantscatteringlorentzviolatinggravity}. This phenomenon has also been examined from thermodynamic and black hole evaporation perspectives~\citep{Bekenstein1998,PhysRevD.7.949,Hawking1975vcx}. If the amplified bosonic field is confined and repeatedly scattered by the black hole, an exponentially growing instability may develop, giving rise to the black hole bomb mechanism~\citep{Press:1972zz,Yoshino_2012,PhysRevLett.109.131102}. Superradiant instabilities have further been proposed as astrophysical probes of ultralight bosonic particles, including axion-like dark-matter candidates~\citep{Brito_2020}. Rotating black holes may also act as natural particle accelerators. Ba\~{n}ados-Silk-West (BSW)~\citep{PhysRevLett.103.111102} demonstrated that the center-of-mass energy of two particles colliding near the horizon of an extremal Kerr black hole can, under critical conditions, become arbitrarily large. This BSW effect has motivated extensive investigations of ultra-high-energy collisions in strong gravitational fields~\citep{Amir_2016,PhysRevD.82.103005,Mukherjee_2019,Pradhan:2014oaa,Liu:2010ja,Mao_2017,Amir:2015pja,Zakria:2015eua,PhysRevD.111.024022,PhysRevD.110.064016,Ovcharenko_2024,Zhang_2025,Rudra:2019ssz,Babar:2021nst}. An important issue that remains is to investigate how the presence of a dilaton parameter alters these energy extraction mechanisms and the corresponding efficiencies. As a result, these mechanisms are expected to undergo notable deviations from their classical counterparts in the Kerr spacetime. The additional coupling introduced by the dilaton field alters the local curvature and frame-dragging behaviour, thereby affecting the efficiency of energy extraction and the dynamics of particle acceleration. A particularly intriguing aspect in this context concerns the range of angular momentum for which the center-of-mass energy of colliding particles diverges, suggesting the possibility of ultra-high-energy interactions in the vicinity of the event horizon of a rotating EMDA black hole. Finally, the dilaton parameter $b$ will be the central focus of the aforementioned study, with a detailed analysis of its role across different energy extraction schemes in the subsequent sections.
	
\vspace{0.2cm}
\noindent
In this work, we have investigated several mechanisms of energy extraction and high-energy particle collision in the spacetime of a rotating EMDA black hole. We first establish the underlying theoretical framework by introducing the rotating EMDA geometry, in which the dilaton hair parameter $b$, satisfying $b\leq 0$, provides a fundamental deviation from the standard Kerr spacetime. The equations governing equatorial particle motion are then derived, thereby elucidating the influence of the parameter $b$ on the geodesic structure of the spacetime. We subsequently examine rotational energy extraction through the Penrose process and show that increasingly negative values of $b$ can significantly enhance the maximum extraction efficiency, particularly in the extremal black hole limit. The irreducible mass and the available rotational-energy reservoir are also evaluated. The results indicate that decreasing $b$ reduces the irreducible mass $M_{\rm irr}$ and correspondingly increases the extractable rotational energy $\mathcal{E}_{\rm rot}$ relative to the Kerr limit. The kinematic conditions governing the production of fragments within the ergoregion are further analysed using the Wald inequality and the Bardeen-Press-Teukolsky inequality. These analyses demonstrate that the lower bounds imposed on the local fragment velocities become progressively less restrictive as the dilaton parameter assumes more negative values. In addition, the superradiant scattering of a minimally coupled test scalar field is investigated. By deriving the associated flux-balance relations, we show that the dilaton parameter modifies the horizon angular velocity $\Omega_H$ and, consequently, alters the frequency interval within which superradiant amplification can occur. Finally, we study the center-of-mass energy $E_{\rm CM}$ of colliding particles near the event horizon. The analysis confirms that the BSW divergence persists in the rotating EMDA spacetime when one of the particles possesses the appropriate critical angular momentum. The parameter $b$ modifies the horizon geometry, the angular velocity of the horizon, and hence the critical angular-momentum threshold required for arbitrarily large collision energies. It should be emphasised that the present analysis is entirely theoretical and is carried out within the test-particle and test-field approximations. Accordingly, the effects of backreaction, self-force, finite-size corrections, quantum-gravitational phenomena, plasma interactions, magnetic fields, accretion flows, radiative losses, and other astrophysical environmental factors are neglected. The results should therefore be interpreted as idealised upper bounds and as indicators of the underlying kinematic properties of the rotating EMDA geometry rather than as direct predictions for realistic astrophysical black holes.
	
\vspace{0.2cm}
\noindent
The paper is organised as follows. Section~\ref{sec_2} introduces the rotating EMDA spacetime, its effective action, metric, horizon and ergoregion structure, extremality condition, and admissible $(a,b)$ parameter space. In section~\ref{sec_3}, the Hamilton--Jacobi equation is separated to obtain the conserved quantities and effective potentials governing geodesic motion. Section~\ref{sec_4} examines the Penrose process and its maximum efficiency in the extremal limit. Subsections~\ref{sec_4_1} and~\ref{sec_4_2} respectively discuss the irreducible mass and extractable rotational energy $\mathcal{E}_{\rm rot}$, and the kinematic bounds on fragment velocities derived from the Wald and Bardeen-Press-Teukolsky inequalities. Section~\ref{sec_5} analyses scalar-field superradiance and derives the associated flux relations and amplification condition. Section~\ref{sec_6} investigates equatorial two-particle collisions and the corresponding center-of-mass energy $E_{\rm cm}$. Finally, section~\ref{sec_7} summarises the principal results and their implications for rotating EMDA black holes. Throughout the paper, we employ rescaled geometrised units with $(G=c=M=1)$, except in sections~\ref{sec_4} and~\ref{sec_5}, where the black hole mass is retained explicitly for clarity.

\section{The Einstein-Maxwell-Dilaton-Axion spacetime}\label{sec_2}
	
The following action describes the four-dimensional EMDA model,
\begin{equation}
		S=\int d^{4} x \sqrt{-g}\left(R-2 \partial_{\mu} \varphi \partial^{\mu} \varphi-\frac{1}{2} e^{4 \varphi} \partial_{\mu} \kappa \partial^{\mu} \kappa-e^{-2 \varphi} F_{\mu \nu} F^{\mu \nu}-\kappa F_{\mu \nu} \check{F}^{\mu \nu}\right), \label{bn}
\end{equation}
	
\vspace{0.2cm}
\noindent
where $R$ denotes the scalar curvature (Ricci scalar), $F_{\mu \nu}$ is the electromagnetic field strength tensor (an antisymmetric rank-2 tensor), and $\check{F}_{\mu \nu}$ represents its Hodge dual. The scalar field $\varphi$ corresponds to the dilaton scalar field, while $\kappa$ denotes the axion field. The metric explicitly for EMDA black hole~\citep{Wei_2013,Flathmann:2015xia} can be obtained by solving the field equations derived from the action~\eqref{bn}.
	
\begin{multline}\label{metric}
	ds^{2}=-\frac{\Delta-a^{2}\sin^{2}\theta}{\Sigma}\,dt^{2}
		+\frac{\Sigma}{\Delta}\,dr^{2}
		+\frac{(\Xi^{2}-a^{2}\Delta\sin^{2}\theta)\sin^{2}\theta}{\Sigma}\,d\phi^{2}+\Sigma\,d\theta^{2}\\[6pt]
		-\frac{2a(\Xi-\Delta)\sin^{2}\theta}{\Sigma}\,dt\,d\phi ,
\end{multline}
	
with the metric functions given by
	
\begin{equation}
		\begin{aligned}
			\Sigma &=r^{2}+2br+a^{2}\cos^{2}\theta,\\[8pt]
			\Delta &=r^{2}-2mr+a^{2},\\[8pt]
			\Xi &=r^{2}+2br+a^{2}. 
		\end{aligned}
\end{equation}
	
	\begin{figure}[h!]
		\centering
		\subfigure[]{\includegraphics[width=7.3cm,height=7.7cm]{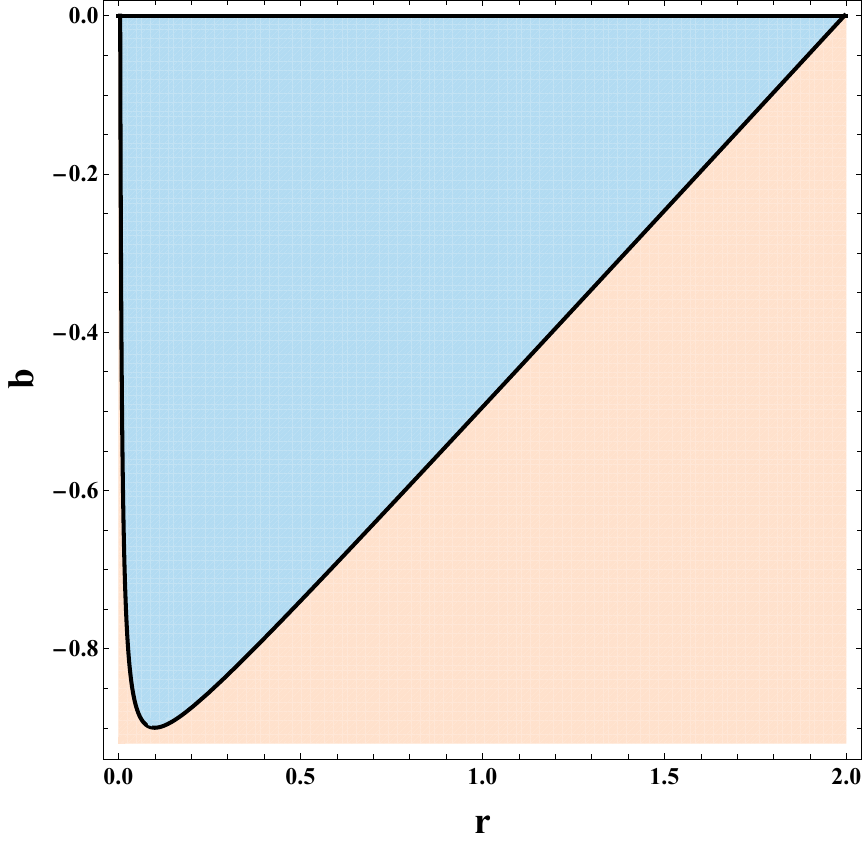}\label{Vf1r_1}} \hspace{0.8cm}
		\subfigure[]{\includegraphics[width=7.3cm,height=7.7cm]{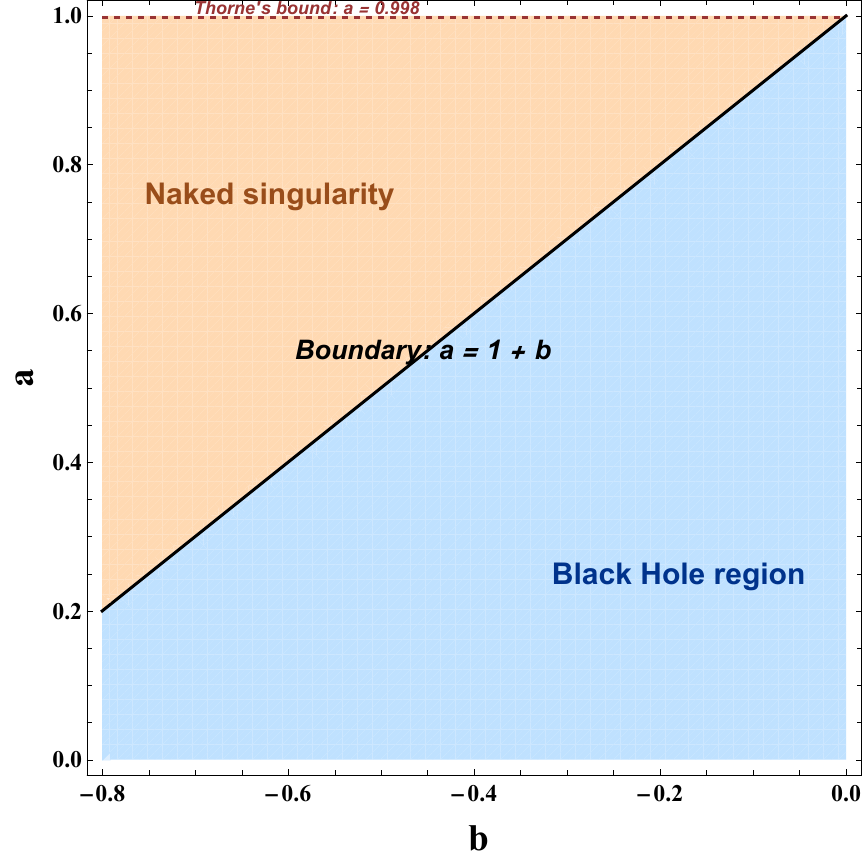}\label{Vf1r_2}}
		\caption{The allowed parameter space of the rotating EMDA black hole for ($M=1$). Panel (a) shows the admissible range of the dilaton parameter $b$, restricted to negative values by the requirement of a real electric charge and further constrained by the positivity of the radial coordinate $r$. Panel (b) presents the $(a,b)$ parameter space, where the shaded region corresponds to black hole solutions bounded by the extremality condition $a=1+b$ (solid curve). The region above this boundary represents naked-singularity configurations. The horizontal dashed line marks Thorne’s astrophysical spin limit, $a=0.998$.}\label{para_space}
	\end{figure}
	
	\vspace{0.2cm}
	\noindent
	The parameter $b$ represents the dilaton charge, while the mass parameter $(m)$ appearing in $\Delta$ is related to the Arnowitt-Deser-Misner (ADM) mass by $M = m-b$ (equivalently $m=M+b$). The parameter $a$ corresponds to the black hole's rotation, with angular momentum given by $J = aM$. The electric charge of the black hole is expressed as $Q = \sqrt{-2\omega b M}$, where $\omega = \exp(2\phi_0)$, $\phi_0$ is the asymptotic value of the dilaton field, see \citep{PhysRevLett.74.1276}. Requiring the electric charge $Q$ to be real and assuming $M > 0$, it follows that the dilaton charge must satisfy $b \leq 0$. In the special case $b = 0$, the solution reduces to the standard Kerr spacetime, recovering the usual vacuum rotating black hole of GR. And in the limiting case $a=0$, the rotating EMDA black hole solution reduces to the non-rotating Garfinkle-Horowitz-Strominger (GHS) dilaton metric \citep{PhysRevLett.74.1276,Ghosh_1994}.
	\begin{table}[h!]
		\centering
		\small
		\renewcommand{\arraystretch}{1.15}
		\caption{The outer and inner horizon radii \(r_{\pm}\) are shown for different values of \(a\) and \(b\), with \(M=1\), together with the horizon separation \(\delta=r_+-r_-\). The table illustrates the EMDA horizon structure, while the extremal energy-extraction analysis is restricted to the regular exterior domain \(-1/3<b\leq0\).}
		\label{Tab_hor}
		\vspace{0.5cm}
		\begin{tabular*}{\textwidth}{@{\extracolsep{\fill}} c c c c c @{}}
			\hline\hline
			$\bf b$ & $\bf a$ & $\bf r_{+}$ & $\bf r_{-}$ & $\bf \delta=r_{+}-r_{-}$ \\
			\hline \hline
			-0.1 & 0.100000 & 1.7944270 & 0.0055728 & 1.7888542 \\
			& 0.300000 & 1.7485280 & 0.0514719 & 1.6970561 \\
			& 0.500000 & 1.6483310 & 0.1516685 & 1.4966625 \\
			& 0.700000 & 1.4656850 & 0.3343146 & 1.1313704 \\
			& 0.900000 & 0.9000000 & 0.9000000 & 0.0000000 \\
			\hline
			-0.2 & 0.100000 & 1.5937250 & 0.0062746 & 1.5874504 \\
			& 0.300000 & 1.5416200 & 0.0583802 & 1.4832398 \\
			& 0.500000 & 1.4245000 & 0.1755002 & 1.2490000 \\
			& 0.700000 & 1.1872980 & 0.4127017 & 0.7745963 \\
			& 0.800000 & 0.8000000 & 0.8000000 & 0.0000000 \\
			\hline
			-0.3 & 0.100000 & 1.3928200 & 0.0071797 & 1.3856403 \\
			& 0.300000 & 1.3324560 & 0.0675445 & 1.2649115 \\
			& 0.500000 & 1.1898980 & 0.2101021 & 0.9797959 \\
			& 0.700000 & 0.7000000 & 0.7000000 & 0.0000000 \\
			\hline
			-0.4 & 0.100000 & 1.1916080 & 0.0083920 & 1.1832160 \\
			& 0.300000 & 1.1196150 & 0.0803848 & 1.0392302 \\
			& 0.500000 & 0.9316625 & 0.2683375 & 0.6633250 \\
			& 0.600000 & 0.6000000 & 0.6000000 & 0.0000000 \\
			\hline
			-0.5 & 0.100000 & 0.9898980 & 0.0101021 & 0.9797959 \\
			& 0.300000 & 0.9000000 & 0.1000000 & 0.8000000 \\
			& 0.500000 & 0.5000000 & 0.5000000 & 0.0000000 \\
			\hline
			-0.6 & 0.100000 & 0.7872983 & 0.0127017 & 0.7745966 \\
			& 0.300000 & 0.6645751 & 0.1354249 & 0.5291502 \\
			& 0.400000 & 0.4000000 & 0.4000000 & 0.0000000 \\
			\hline
			-0.7 & 0.100000 & 0.5828427 & 0.0171573 & 0.5656854 \\
			& 0.300000 & 0.3000000 & 0.3000000 & 0.0000000 \\
			\hline
			-0.8 & 0.100000 & 0.3732051 & 0.0267949 & 0.3464102 \\
			& 0.200000 & 0.2000000 & 0.2000000 & 0.0000000 \\
			\hline\hline
		\end{tabular*}
	\end{table}
	
	\vspace{0.2cm}
	\noindent
	The structure of the horizons is determined by solving $(\Delta=0)$. Using ($m=M+b$), the horizon radii can be written in terms of the ADM mass $M$ as
	
	\begin{equation}
		r_\pm = (M+b) \pm \sqrt{\left( M+b \right)^2 -a^2} \, .
	\end{equation}
	
	\vspace{0.2cm}
	\noindent
	Comprehensively, the EMDA spacetime (\ref{bn}) admits two horizons, obtained from the quadratic condition $\Delta=0$. These are identified as the Cauchy horizon ($r_-$) and the event horizon ($r_+$). Unlike gauged supergravity black holes, no cosmological horizons arise here since the EMDA geometry is asymptotically flat. The outer horizon, often referred to as the surface of no return, represents the boundary beyond which no information or matter can escape to infinity. The extremal EMDA black hole corresponds to the case when the two roots of $\Delta=0$ coincide, i.e. $r_- = r_+ = r_E$, where $r_E$ denotes the degenerate horizon. This condition fixes a precise relation between the rotation parameter $a$ and the dilaton parameter $b$ (or equivalently, the ADM mass). For sub-extremal configurations ($r_+ > r_-$), the spacetime represents a regular black hole, whereas for $a^2 > (M+b)^2$ the horizons disappear and the solution reduces to a naked singularity. Thus, the transition from non-extremal to extremal, and finally to a naked singularity, is governed by the tuning of $a$ and $b$. Because of the intricate interplay between these parameters, the horizon structure is best explored numerically. A useful diagnostic is the horizon separation parameter $(\delta = r_+ - r_-)$, which quantifies the distance between the event horizon and the Cauchy horizon. As shown in our numerical results (see Tables~\ref{Tab_hor}), $\delta$ decreases as the black hole approaches the extremal limit and vanishes when $r_+=r_-$. For astrophysical black holes, the spin parameter is subject to an upper limit, commonly referred to as Thorne's bound~\citep{1974ApJ...191..507T}, which restricts $a_{E^{*}} \leq 0.998$. To ensure that the horizon radii $r_{\pm}$ are real, the condition $a^2 \leq (M + b)^2$ must be satisfied (see Table~\ref{Tab_hor}). 
	
	\vspace{0.2cm}
	\noindent
	It is worth emphasising that the parameter $b$ should not be regarded as an independent arbitrary deformation parameter once the physical charge and the asymptotic value of the dilaton field are specified. Our primary motivation is to investigate how the inclusion of $b$ (theoretical parameter) modifies the energy-extraction and particle-acceleration mechanisms relative to Kerr, with particular emphasis on the negative $b$ domain. For astrophysical black holes, which are expected to remain nearly electrically neutral owing to efficient plasma-neutralization processes, the dilaton parameter is therefore constrained to $b\simeq 0$, in which limit the EMDA solution reduces to the Kerr geometry. Fig.~(\ref{para_space}) shows the admissible parameter space of the rotating EMDA black hole. Panel (\ref{Vf1r_1}) shows the range of the dilaton parameter $b$ corresponding to positive values of the radial coordinate $r$, with the shaded region denoting the admissible domain adopted throughout the present study to analyse the various energy-extraction schemes and the BSW effect. Panel (\ref{Vf1r_2}) shows the $(b,\:a)$ plane and separates the black hole and naked singularity regions. The diagonal curve labelled 'Boundary: $a=1+b$' gives the extremality limit with $(M=1)$. Below this curve, spacetime admits horizons (the black hole region). Above it, the horizons disappear, and the solution corresponds to a naked singularity. The same panel also includes Thorne's bound: $a=0.998$. This provides an astrophysical upper limit on the spin and indicates the practically relevant region of the black hole. Thus, the parameter-space plots should be read as an exploratory scan of the model, not as an observationally established range. Unlike the Kerr black hole, the presence of the dilaton parameter $b$ permits scenarios in which both $r_+$ and $r_-$ can assume negative values. Here, we focus our analysis exclusively on cases where the horizon radii are positive. 
	
	\vspace{0.2cm}
	\noindent
	Recent EHT based studies have constrained the dilaton charge of the EMDA like Kerr--Sen spacetime, although the parameter conventions used in these works differ from that adopted in the present analysis. In terms of the dilaton parameter $r_2$, the M87$^{*}$ shadow favours the Kerr scenario and excludes $r_2>0.48$, whereas Sgr A$^{*}$ shows a marginal preference towards the Kerr--Sen case, while the Kerr limit remains allowed, with $r_2>1$ excluded~\citep{Sahoo2025}. Earlier EHT based analyses also reported preferred non-zero ranges of $r_2$ for particular shadow measurements, while retaining the Kerr limit within the observational uncertainties~\citep{Sahoo:2023czj}. Independent constraints have also been obtained from the orbit of the S2 star. Using the convention $b_{\rm KS}=r_2/2$, Fern\'andez et al. obtain $b_{\rm KS}\lesssim1.4M$ at $95\%$ confidence after incorporating the measured relativistic orbital precession of S2, while mock future GRAVITY-quality observations forecast $b_{\rm KS}\lesssim0.033M$~\citep{Fernandez:2023kro}. Since these studies employ parameter conventions different from that adopted here, we quote the observational constraints in their original notation. Accordingly, the negative values of $b$ considered in the present work should be regarded as illustrative theoretical choices rather than observationally inferred values.
	
	\vspace{0.2cm}
	\noindent
	The ergoregion is defined by $g_{tt}=0$, i.e.\ $(\Delta-a^{2}\sin^{2}\!\theta=0)$, which in the equatorial plane yields
	$r_{\rm ergo}=2m=2(M+b)$. Inside it, frame dragging enforces co-rotation, forbidding retrograde motion. The curvature singularity occurs where $\Sigma=r^{2}+2br+a^{2}\cos^{2}\!\theta=0$, at which the Kretschmann scalar diverges. In the equatorial plane $(\theta=\tfrac{\pi}{2})$ this condition reduces to $r(r+2b)=0$, giving $r=0$ and $r=-2b$. For $b<0$, the second root lies at positive $r$, indicating that the singular set is not confined to $r\le 0$ in this parametrisation. Its geometry can vary from a deformed torus to two closed surfaces depending on $(a,b)$~\citep{flathmann2016analytic}. Only geodesics reaching $r=0,\theta=\tfrac{\pi}{2}$ terminate at the singularity; other trajectories may extend across the $r=0$ surface into the $r<0$ region, where the effective gravitational interaction can become repulsive.
		
	\section{Equations of Motion for EMDA Black Hole}\label{sec_3}
	
	The geodesic structure of this spacetime has been extensively analysed in earlier works \citep{Wei_2013,Flathmann:2015xia}. For completeness, we briefly summarise the relevant features here, beginning with the Hamilton-Jacobi equation that governs particle motion in this black hole geometry,
	
	\begin{equation}
		\frac{\partial W}{\partial\alpha}
		=-\frac{1}{2}g^{\mu\nu}\frac{\partial W}{\partial x^{\mu}}
		\frac{\partial W}{\partial x^{\nu}},\label{jacobiequation}
	\end{equation}
	
	\vspace{0.2cm}
	\noindent
	where $\alpha$ is an affine parameter along the geodesics and $W$ is the Jacobi action. For this black hole spacetime~\eqref{metric}, the Jacobi action $W$ can be separated as
	
	\begin{equation}
		W=\frac{1}{2}m^{2}_{0}\alpha+L\phi+W_{r}(r)+W_{\theta}(\theta)-Et,\label{action}
	\end{equation}
	
	where $W_{r}$ and $W_{\theta}$ are functions of $r$ and $\theta$, respectively. The parameters $m_{0}$, $E$, and $L$ denote the test particle's mass, energy, and angular momentum, respectively. Therefore, substituting~\eqref{action} into~\eqref{jacobiequation}, we obtain the geodesic equation,
	
	\begin{equation}
		\frac{dr}{d\tau} = \pm \sqrt{R}, \label{eq:radial}
	\end{equation}
	
	\begin{equation}    
		\frac{d\theta}{d\tau} = \pm \sqrt{\Theta}, \label{eq:theta}
	\end{equation}
	
	\begin{equation}
		\frac{d\phi}{d\tau}=\frac{a(a^{2}E-aL+E(r^{2}+2br))}{\Sigma}
		+L\csc^{2}\theta-aE,\label{tr_1}
	\end{equation}
	
	\begin{equation}
		\frac{dt}{d\tau}=\frac{\Xi(a^{2}E-aL+r(r+2b)E)}{\Sigma}
		+aL-a^{2}E\sin^{2}\theta,\label{tr_2}
	\end{equation}
	
	where
	
	\begin{equation}\label{radial_1}
		\begin{aligned}
			R(r) &= (\Xi E - aL)^2
			- \Delta\!\left[\,Q + (aE - L)^2 + m_0^2\,(r^2 + 2br)\,\right],\\[10pt]
			\Theta(\theta) &= Q
			- \cos^2\theta\left[(m_0^2 - E^2)a^2 - \frac{L^2}{\sin^2\theta}\right].
		\end{aligned}
	\end{equation}
	
	\vspace{0.2cm}
	\noindent
	Following~\citep{Flathmann:2015xia}, we adopt the Mino time parametrisation~\citep{PhysRevD.67.084027} in deriving Eqs.~\eqref{eq:radial}--\eqref{tr_2}. In our notation, $\tau$ represents the Mino time, while $\alpha$ denotes the proper time parameter, with the two related by $\Sigma d\tau = d\alpha$. Accordingly, all derivatives in Eqs.~\eqref{eq:radial}--\eqref{tr_2} are to be understood as derivatives with respect to the Mino parameter $\tau$. This reparametrisation is used only to simplify the separated geodesic equations, while the conserved energy $E$ and angular momentum $L$ retain their standard definitions associated with the stationarity and axial symmetry of the spacetime. Here, $Q$ is a constant of separation, and if $b=0$, it is just the geodesic for the Kerr black hole~\citep{chandrasekhar1983mathematical}.
	
	\section{Energy Extraction in EMDA Black Holes: The Penrose Process}\label{sec_4}
	
	Following the discovery of the ergosphere in rotating black holes, Roger Penrose proposed a gedankenexperiment~\citep{PhysRevLett.14.57} to demonstrate that the energy of a particle within the ergoregion, as measured by an observer at infinity, can become negative. He further argued that if a test particle enters the ergosphere and decays at a turning point in its geodesic trajectory into two identical fragments, one with positive energy may escape to infinity. In contrast, the other, carrying negative-energy, is absorbed back by the black hole. In principle, this realisation leads to a unique way of extracting rotational energy out of a rotating black hole, hence naming it as the Penrose process~\citep{Penrose1971uk,Mukherjee_2019,PhysRevD.89.024023,1985JApA,Rudra:2019ssz,1986ApJ,ganguly2014,PhysRevD8404,Abdujabbarov_2011,Chen:2013vja,Ortiqboev:2026mip,Ashraf:2025lxs,Ortiqboev:2025jzu,Ortiqboev:2024mtk}. Here, we will discuss the workability of energy extraction for an EMDA black hole and the original Penrose process, along with the further bounds from Wald and the Bardeen-Press-Teukolsky inequality. As discussed previously, in the Penrose process, a massive particle entering the ergoregion reaches a turning point ($\dot r = 0$) and subsequently splits into two massless fragments. From Eq.~(\ref{radial_1}), we have, $R = 0$ which leads to
	
	\vspace{0.2cm}
	\begin{equation}
		E=\frac{\big(\Xi-\Delta\big)aL\pm \sqrt{\Delta}\sqrt{\big(a^2-\Xi\big)^2L^2+\big(\Xi^2-\Delta a^2\big)\big(Q+m^2_{0}(r^2+2br)\big)}}{\big(\Xi^2-\Delta a^2\big)},\label{sup13}
	\end{equation}
	
	\vspace{0.60cm}
	\begin{equation}
		L=\frac{\big(\Xi-\Delta\big)aE\pm \sqrt{\Delta}\sqrt{\big(\Xi-a^2\big)^2E^2+\big(a^2-\Delta\big)\big(Q+m^2_{0}(r^2+2br)\big)}}{\big(a^2-\Delta\big)}. \label{sup14}
	\end{equation}
	
	\vspace{0.2cm}
	\noindent
	The $\pm$ signs in the above equations correspond to co-rotating and counter-rotating orbits, respectively. 
	Without loss of generality, for a unit mass particle initially at rest at infinity, the energy may be normalized as $E=1$. We now use Eq.~\eqref{sup13} to determine the conditions under which a particle can occupy a negative-energy state, as measured by an observer at infinity. Following the standard Chandrasekhar's treatment~\citep{chandrasekhar1983mathematical}, we derive (see Appendix~\ref{app:append_a}) the following fundamental identity for the spacetime under consideration:
	
	\begin{equation}
		(\Xi^2-\Delta a^2) (\Delta-a^2) = r^2(r-2M-2b)\Big[r(r+2b)^2+a^2(r+2M+6b)\Big]. \label{inden}
	\end{equation}
	
	\vspace{0.2cm}
	\noindent
	This identity reduces to the corresponding Kerr factorisation when $b=0$. In the present scenario, only the positive sign on the right-hand side of Eq. (\ref{sup13}) is taken into account. Therefore, for the existence of negative-energy states, i.e., $E < 0$, it is necessary that $L < 0$ together with the satisfaction of the following criterion. Further constraints may then be obtained by simplifying the resulting inequality.
		
	\begin{equation}
		\big(\Xi-\Delta\big)^2 a^{2}L^{2} > \Delta\Bigg(\big(a^2-\Xi\big)^2L^2+\big(\Xi^2-\Delta a^2\big)\big(Q+m^2_{0}(r^2+2br)\big)\Bigg). \label{in_1}
	\end{equation}
		
	\vspace{0.2cm}
	\noindent
	Using the factorisation derived in Appendix~(\ref{app:append_a}) together with Eq.~\eqref{inden}, Eq.~\eqref{in_1} reduces to
		\begin{equation}
		r(r-2M-2b)L^2 + \Delta\left[Q + m_0^2(r^2+2br)\right] < 0. \label{in_2}
	\end{equation}
	
	\vspace{0.2cm}
	\noindent
	From inequality~\eqref{in_2}, it follows that the following condition must be satisfied:
	
	\begin{equation}
		E<0\Longleftrightarrow L<0\quad\mathrm{and}\quad \big(r-2M-2b\big) < -\frac{\Delta}{rL^2}\left[Q + m_0^2(r^2+2br)\right].
	\end{equation}
	
	\vspace{0.2cm}
	\noindent
	For $b=0$, the above inequality reduces to the corresponding expression for the Kerr black hole~\citep{chandrasekhar1983mathematical}. The relation further indicates that negative-energy states are possible only for particles with negative angular momentum, provided that their radial position within the ergoregion also satisfies the above inequality.
	
	\vspace{0.2cm}
	\noindent
	Let us consider an incident massive particle of unit rest mass $(m_{0}=1)$, carrying conserved energy $(E^{(c)}>0)$ and angular momentum $(L^{(c)})$, that enters the ergoregion of the EMDA spacetime while moving on the equatorial plane $(\theta=\pi/2, Q=0)$. Let the particle be broken into two massless ($m_1=m_2=0$) pieces with energies and angular momenta $(E^{(a)}, L^{(a)})$ and , $(E^{(b)}, L^{(b)})$ respectively, one crossing the black hole horizon while the other one comes out of the ergoregion. From Eq.~\eqref{sup13}, we note that to ensure positive energy in the limit $a \to 0$, only the positive sign must be retained. On the other hand, for $a \neq 0$, a necessary condition for the existence of negative-energy states is $L < 0$, corresponding to counter-rotating orbits with respect to the black hole's angular momentum. We have from (\ref{sup14})
	
	\begin{equation}
		\centering
		\begin{aligned} 
			L^{(c)} &=  \frac{\left(\Xi - \Delta\right)\:a + \sqrt{\Delta}\,\sqrt{\left(\Xi - a^2\right)^2 + \left(a^2 - \Delta\right)\left(r^2 + 2br\right)}}{\left(a^2 - \Delta\right)}E^{(c)} 
			=  E^{(c)}\: \alpha^{(c)}, \\[10pt]
			L^{(a)} &=  \frac{\left(\Xi - \Delta\right)\:a - \sqrt{\Delta}\left(\Xi - a^2\right)}{\left(a^2 - \Delta\right)}E^{(a)} 
			= E^{(a)}\: \alpha^{(a)}, \\[10pt]
			L^{(b)} &=  \frac{\left(\Xi - \Delta\right)\:a + \sqrt{\Delta}\left(\Xi - a^2\right)}{\left(a^2 - \Delta\right)} E^{(b)}
			= E^{(b)}\: \alpha^{(b)}.
		\end{aligned} \label{sup15}
	\end{equation}
	
	\vspace{0.2cm}
	Using the above equations, the stationary-axisymmetric conservation laws  read
	
	\begin{equation}
		E^{(c)}=E^{(b)}+E^{(a)}, \qquad L^{(b)}+L^{(a)}= E^{(a)}\alpha^{(a)}+E^{(b)}\alpha^{(b)} = E^{(c)}\alpha^{(c)}=L^{(c)},\label{sup16}
	\end{equation}
	
	\vspace{0.2cm}
	which can be solved to get the energies
	
	\begin{equation}
		E^{(a)}=\frac{\alpha^{(c)}(r, E^{(c)})-\alpha^{(b)}(r)}{\alpha^{(a)}(r)-\alpha^{(b)}(r)} E^{(c)}, \qquad E^{(b)}=\frac{\alpha^{(a)}(r)-\alpha^{(c)}(r,E^{(c)})}{\alpha^{(a)}(r)-\alpha^{(b)}(r)}E^{(c)}.\label{sup17}
	\end{equation}
	
	\vspace{0.2cm}
	We find using (\ref{sup15}) we get
	
	\begin{equation}
		\begin{aligned}
			E^{(a)} &= -\frac{1}{2}
			\left[
			\left(
			1 + \frac{(a^{2}-\Delta)\,(r^{2}+2br)}
			{(\Xi-a^{2})^{2}\,(E^{(c)})^{2}}
			\right)^{1/2}
			- 1
			\right] E^{(c)}, \\[12pt]
			E^{(b)} &= \frac{1}{2}
			\left[
			\left(
			1 + \frac{(a^{2}-\Delta)\,(r^{2}+2br)}
			{(\Xi-a^{2})^{2}\,(E^{(c)})^{2}}
			\right)^{1/2}
			+ 1
			\right] E^{(c)} .
		\end{aligned}
		\label{sup18}
	\end{equation}
	
	\vspace{0.2cm}
	Thus, it is clear once again that energy extraction, i.e., $E^{(b)}>E^{(c)}$ would be possible when we are inside the ergosphere (\ref{sup13}). The amount of energy extracted is given by 
	
	\begin{equation}
		\Delta E =E^{(b)}-E^{(c)} =\frac12 \left[ \sqrt{\left(1+\frac{\left(r^2+2br\right)\left(a^2-\Delta\right)}{(\Xi-a^2)^{2}(E^{(c)})^2} \right)}-1\right]E^{(c)}.\label{sup19}
	\end{equation}
	
	\vspace{0.2cm}
	Consequently, the maximum of the energy extracted would correspond to when the incident particle splits at the horizon $r_H$, i.e. $\Delta=0$, and we have
	
	\begin{equation}
		\left. \Delta E \right|_{\max}= \frac12 \left[-1+ \sqrt{\left(1+\frac{a^2}{(E^{(c)})^2(r^2_{H}+2br_{H})} \right)}\right]E^{(c)}.	\label{sup20}
	\end{equation}
	
	\vspace{0.2cm}
	Therefore, at the extremal limit ($a=a_{E}$, $r=r_{E}$), the maximum efficiency can be expressed as
	
	\begin{equation}
		\eta_{\max}= \frac12 \left[ \sqrt{\left(1+\frac{a_{E}^2}{(r^2_{E}+2br_{E})} \right)}-1\right].\label{sup21}
	\end{equation}
	
	\vspace{0.2cm}
	It is customary to set the reference energy as $E^{(c)} = 1$, which serves as the normalisation condition for the analysis. It is noteworthy that this result depends explicitly on the parameters $a$, $b$ and $r_{E}$. The expression retains the same formal structure as in the Kerr black hole, with the dilaton parameter $b$ contributing implicitly via its modification of $r_{E}$.  In the limiting case $b=0$, the standard extremal Kerr result is recovered with $\eta_{\max}=0.207$ \citep{chandrasekhar1983mathematical}. Since the Penrose process considered here represents an idealised reversible upper bound, and the numerical values are tied to the imposed extremal condition $a_E=r_E=1+b$, we report the corresponding efficiencies to four significant figures (see Table~\ref{tab_effi}). It is important to emphasize that Eq. \eqref{sup21} represents an ideal geodesic upper bound for the Penrose process. The derivation assumes a local splitting of an incident massive particle, normalized by ($E^{(c)}=1)$, into two on-shell fragments in the equatorial plane. In the present calculation the fragments are taken to be massless, and therefore satisfy $(p_{(i)}^\mu p^{(i)}_\mu=0)$, while the timelike normalization $(u^\mu u_\mu=-1)$ applies to the parent massive particle. The negative-energy fragment arises because the Killing vector (($\partial_t)^\mu$) becomes spacelike inside the ergoregion, so the conserved Killing energy ($E=-p_t$) can be negative without violating the on-shell condition. For a future-directed fragment crossing the horizon, the relevant condition is $(E-\Omega_H L\geq 0)$, which is compatible with (E<0) when the fragment has sufficiently negative angular momentum. Thus (E<0) is not an additional ad hoc assumption, but follows from the allowed geodesic branch and the local conservation of energy and angular momentum at the splitting point.
	
	\vspace{0.2cm}
	\begin{table}[h!]
		\centering
		\begin{tabular}{|r|c|c|c|c|}
			\hline
			$\bf b$ & $\bf a_{E}$ & $\bf r_{E}$ & $\bf \eta_{\max}$ & $\bf \eta_{\max}$ (\%) \\
			\hline \hline
			0.0  & 1.0 & 1.0 & 0.2071 & 20.7106 \\
			-0.1 & 0.9 & 0.9 & 0.2559 & 25.5928 \\
			-0.2 & 0.8 & 0.8 & 0.3660 & 36.6025 \\
			-0.3 & 0.7 & 0.7 & 0.9142 & 91.4213 \\
			\hline
		\end{tabular}
		\caption{Maximum efficiency $\eta_{\max}$ (at extremality) computed in view of the variation of $b$ with $a_E$ and $r_{E}$.}\label{tab_effi}
	\end{table}
	
	\begin{figure}[h!]
		\centering
		\subfigure[]{\includegraphics[width=7.5cm,height=8.2cm]{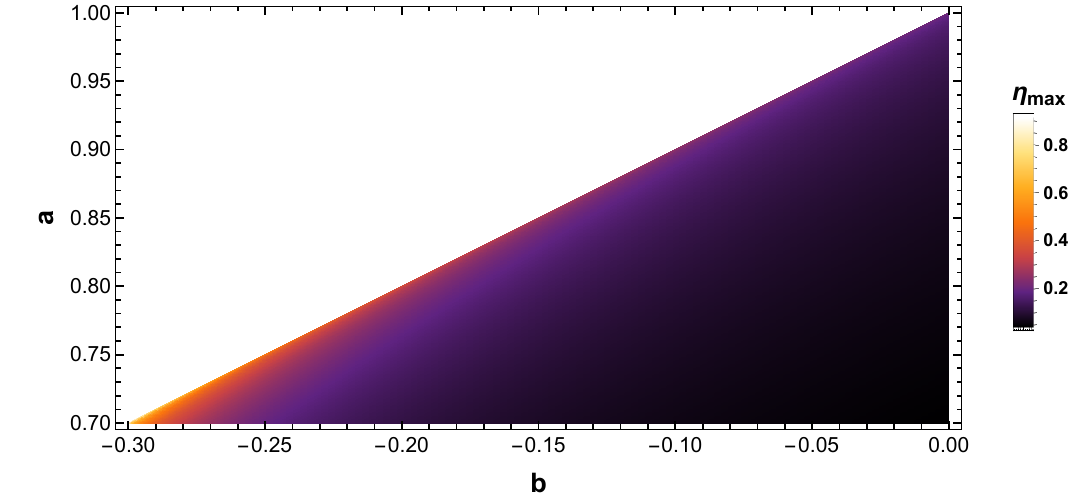}\label{effi_1}} \hspace{0.8cm}
		\subfigure[]{\includegraphics[width=7.3cm,height=7.9cm]{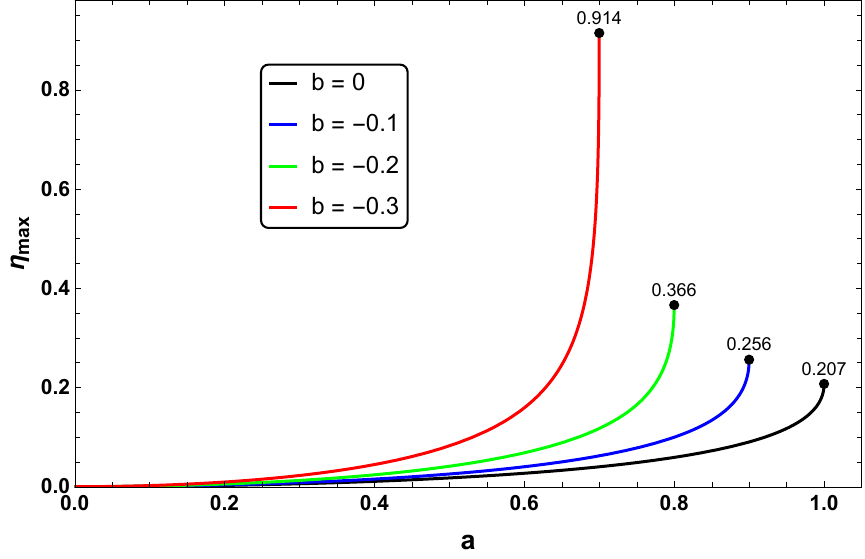}\label{effi_2}}
		\caption{The efficiency of the Penrose process as a function of the parameters $b$ and $a$ is shown. The left panel displays the density (contour) distribution of $\eta_{\max}$ in the $(b,a)$ plane, while the right panel shows $\eta_{\max}$ versus $a$ for different values of $b$.}\label{effi_plot_w}
	\end{figure}
	
	\vspace{0.2cm}
	\noindent
	Compared to the Kerr black hole, an EMDA black hole with a negative dilaton parameter $b$ can yield a substantially larger ideal Penrose process efficiency near extremality, although the allowed extremal spin decreases according to $a_E=r_E=1+b$. The enhancement is therefore not controlled by the extremal spin alone. Rather, the relevant factor entering the square-root term in Eq.~\eqref{sup21} is $\frac{a_E^2}{r_E^2+2br_E}$. Using $a_E=r_E=1+b$, this factor becomes $\frac{a_E^2}{r_E^2+2br_E}=\frac{1+b}{1+3b}$. Thus, although $a_E$ decreases linearly as $b$ becomes more negative, the geometric denominator $1+3b$ decreases more rapidly within the regular exterior domain $-1/3<b\leq0$. This overcompensates the reduction in the available extremal spin and produces the rapid increase in the ideal energy-extraction efficiency.
	
	\vspace{0.2cm}
	\noindent
	In comparison with the standard Kerr limit and other Kerr-family black hole benchmarks discussed in the literature~\citep{Abbasi:2025,chandrasekhar1983mathematical,fatima2025revisitingpenroseprocessrotating,Fatima:2025idf,Mukherjee_2019,Rudra:2019ssz,Kar:2025anc,Pradhan_2019}, the present EMDA case exhibits a substantially larger formal upper bound within the same test-particle approximation. For instance, the extremal Kerr value is recovered at $b=0$, giving $\eta_{\max}\simeq20.71\%$, whereas the illustrative extremal EMDA case with $b=-0.3$ gives
	$\eta_{\max}\simeq91.42\%$ (see Table~\ref{tab_effi}). This enhancement should be understood as a model-dependent theoretical upper bound within the test-particle and reversible horizon-process approximation, rather than as a universal maximum over all rotating black hole geometries. Moreover, the result should not be extrapolated to the limit $b\to-1$. The regular exterior condition requires the extremal horizon to remain outside the positive equatorial singular surface $r=-2b$, which gives $r_E>-2b \Longrightarrow b>-\frac{1}{3}$. Therefore, the large efficiency obtained near $b=-0.3$ should be interpreted as a near-boundary effect within the admissible EMDA domain. In realistic astrophysical environments, backreaction, radiative losses, plasma effects, magnetic fields, and finite-size corrections are expected to reduce the attainable efficiency. Hence the values reported here should be regarded as idealised theoretical benchmarks rather than direct astrophysical predictions.
	
	\vspace{0.2cm}
	A careful assessment of the parameter space requires verifying that the curvature singularity remains cloaked by the event horizon for all configurations analysed in the present work. The curvature invariants, such as the Kretschmann scalar, diverge when $\Sigma=r^2+2br+a^2\cos^2\theta=0$. On the equatorial plane, $\theta=\pi/2$, this condition reduces to $\Sigma_{\rm eq}=r^2+2br=r(r+2b)=0$, which gives $r=0$, $r_{\rm sing}=-2b$. For $b<0$, the second root lies at a positive radial coordinate. In the extremal EMDA branch, with $M=1$, the horizon is located at $r_E=a_E=1+b$. Thus, the condition that the positive equatorial singularity remains hidden inside the extremal horizon is $r_{\rm sing}<r_E$ . Using $r_{\rm sing}=-2b$ and $r_E=1+b$, this condition becomes $-2b<1+b \Longrightarrow b>-\frac{1}{3}$. Therefore, the bound $b>-1/3$ corresponds to the requirement that the positive equatorial curvature singularity remains inside the event horizon. For the illustrative value $b=-0.3$, one obtains $r_{\rm sing}=0.6, r_E=0.7$, so that the singularity is still cloaked by the horizon, with $r_E-r_{\rm sing}=0.1$ for $M=1$ normalisation. At the limiting value $b=-1/3$, the equatorial singularity coincides with the degenerate horizon. Hence this endpoint represents a singular limiting configuration rather than a regular extremal black hole. For $b<-1/3$, the singularity would lie outside the horizon, leading to a naked-singularity regime. Accordingly, the configurations analysed in this work are restricted to the physically admissible regular exterior domain $-\frac{1}{3}<b\leq0$.
	
	\vspace{0.2cm}
	\noindent
	Finally, It should be stressed that the large Penrose efficiencies reported here, particularly the $\sim 91\%$ value for $b = -0.3$, are obtained within the test-particle approximation and represent idealised reversible upper bounds. The capture of a negative-energy fragment with energy $\sim 0.91E_{\rm in}$ would constitute a non-perturbative change to the black hole mass and angular momentum, driving it away from the extremal configuration that enables the large efficiency. The black hole would evolve towards a sub-extremal state with a larger horizon area, in accordance with the second law of black hole thermodynamics, and a correspondingly reduced extractable energy reservoir. A complete treatment of the back-reaction would require solving the full EMDA field equations with a dynamical matter source, which is beyond the scope of this work. The efficiency figures should therefore be interpreted as indicators of the kinematic potential of the EMDA geometry rather than as predictions for realistic astrophysical processes. This caveat applies equally to the superradiance and BSW results obtained in subsequent sections, all of which are derived within the test-field and test-particle limits.
	
	\subsection{\label{ree} The Irreducible Mass \&  Rotational Extraction of Energy}\label{sec_4_1}
	
	Through seminal contributions, Christodoulou~\citep{PhysRevLett.25.1596,PhysRevD.4.3552}
	derived a fundamental relation connecting the total energy of a charged, rotating black hole with its irreducible mass. This result, now considered a cornerstone in black hole physics, provides deep insight into the fundamental limits of energy extraction processes from such compact objects. It is by now a well-acknowledged principle that the surface area of a black hole event horizon is non-decreasing, a fact that underpins much of the thermodynamical interpretation of black hole mechanics. The extraction of energy from a rotating black hole results in a decrease in both its mass $M$  and angular momentum $L$, leaving behind what is known as the irreducible mass, $M_{irr}$, of the black hole. The total mass of a black hole, also referred to as the asymptotic mass, is dependent on the observer's location. According to the newly proposed horizon mass theorem \citep{ha2007newtheoremblackholes}, the mass at the event horizon is always twice the irreducible mass as observed from spatial infinity. Moreover, the irreducible mass constitutes a fundamental quantity that delineates the theoretical upper bound on the extractable energy from a rotating black hole. Motivated by this remarkable framework, in the present section, we aim to extend the analogy and calculate the corresponding irreducible mass for the EMDA black hole. The irreducible mass not only reflects the intrinsic geometry of the event horizon but also integrates modifications arising from the presence of the dilaton parameter $b$. These contributions lead to significant corrections in the theoretical limits of energy extraction mechanisms, notably influencing the efficiency of the Penrose process.
	
	\vspace{0.2cm}
	Therefore, the irreducible mass can be defined as the component of the black hole's mass~\citep{Carroll:2004st}
	
	\begin{equation}
		{M}_{irr} = \sqrt{\frac{{\cal A}}{16\pi G^2}}. \label{mi}
	\end{equation}
	
	\vspace{0.2cm}
	The horizon area of a rotating black hole is calculated by performing the surface integral over the angular coordinates on the two-dimensional horizon cross-section at a fixed radius $r=r_{+}$~\citep{PhysRevLett.25.1596}.
	
	\begin{equation}
		\left.  {\cal A}=\int_0^{2\pi}d\phi\int_0^\pi d\theta \sqrt{g_{\theta \theta}\: g_{\phi \phi}}\right|_{r_{+}},\label{area_1}
	\end{equation}
	
	where
	
	\begin{equation}
		\left.  \sqrt{g_{\theta \theta}\: g_{\phi \phi}}\right|_{r_{+}} = \left. \sqrt{\Sigma\cdot \frac{\Xi^2}{\Sigma}\sin^2\theta} \right|_{r_{+}} = (r_{+}^{2}+2br_{+}+a^{2})\sin\theta. \label{area_2}
	\end{equation}
	
	\vspace{0.2cm}
	Substituting the above equation into Eq.~\eqref{area_1} we get
	
	\begin{equation}
		\begin{aligned}
			{\cal A}&=\int_0^{2\pi}\int_0^\pi (r_{+}^{2}+2br_{+}+a^{2})\sin\theta \\[10pt]
			&=8\pi(1+2b)r_{+}.\label{area_3}
		\end{aligned}
	\end{equation}
	
	\vspace{0.2cm}
	Putting the above results in Eq.~\eqref{mi}, the irreducible mass for EMDA black hole takes the form
	
	\begin{equation}
		{M}_{irr} = \sqrt{\frac{1}{2}(1+2b)\,r_{+}}. \label{mi1}
	\end{equation}
	
	\vspace{0.2cm}
	Consequently, the maximum extractable energy from an extremal EMDA black hole is given by
	
	\begin{equation}
		\mathcal{E}_{rot} = M - M_{irr} = 1 - \sqrt{\tfrac12(1+2b)\,\Big((1+b) + \sqrt{\left( 1+b \right)^2 -a^2}\Big)}.
	\end{equation}
	
	\vspace{0.2cm}
	\begin{table}[h!]
		\centering
		\begin{tabular}{|c|c|c|c|c|c|}
			\hline
			$\bf b$ & $\bf a_E$ & $\bf r_E$ & $\bf M_{\rm irr}$ & $\bf \mathcal{E}_{\rm rot}$ & $\bf \mathcal{E}_{\rm rot}(\%)$ \\
			\hline\hline
			\phantom{-}0.0 & 1.0000000 & 1.0000000 & 0.7071068 & 0.2928932 & 29.289322 \\
			-0.1 & 0.9000000 & 0.9000000 & 0.6000000 & 0.4000000 & 40.000000 \\
			-0.2 & 0.8000000 & 0.8000000 & 0.4898979 & 0.5101021 & 51.010205 \\
			-0.3 & 0.7000000 & 0.7000000 & 0.3741657 & 0.6258343 & 62.583426 \\
			\hline
		\end{tabular}
		\caption{The net extractable rotational energy from an extremal EMDA black hole for different values of the dilaton parameter $b$ is shown, where $M_{irr}$ is the irreducible mass of the black hole.}\label{irr_tab}
		\label{tab_Erot_ext}
	\end{table}
	
	\vspace{0.2cm}
	\noindent
	Table~(\ref{irr_tab}) summarises how the dilaton parameter $b$ reshapes the reversible energy budget of the EMDA black hole. Since the horizon area fixes the irreducible mass, $M_{\rm irr}=\sqrt{\mathcal{A}/(16\pi G^{2})}$, the reduction of the horizon size along the extremal EMDA family as $b$ becomes more negative leads to a monotonic decrease of $M_{\rm irr}$ (Table~\ref{irr_tab}). 
	It is interesting to note that the positivity of the horizon area imposes an	additional restriction on the allowed range of the dilaton parameter. For the EMDA black hole, the horizon area is given by
	$A_H=4\pi\Xi_H$, where
	$\Xi_H=r_H^2+2br_H+a^2$. Using the horizon condition
	$\Delta(r_H)=0$, one obtains
	$a^2=2(M+b)r_H-r_H^2$, and hence
	$\Xi_H=2(M+2b)r_H$. This becomes
	$\Xi_H=2(1+2b)r_H$ for $M=1$. Since the physical outer horizon satisfies
	$r_H>0$, the requirement $A_H>0$ implies $1+2b>0$, or
	$b>-1/2$. Thus, for $M=1$, the area-positive domain becomes
	$-1/2<b\leq0$. The monotonic decrease of $M_{\rm irr}$ with increasingly
	negative $b$ should therefore be interpreted within this area-positive
	domain. A stronger restriction arises when the extremal horizon is further
	required to remain outside the positive equatorial zero of the metric function
	$\Sigma$. In the EMDA geometry,
	$\Sigma=r^2+2br+a^2\cos^2\theta$, which on the equatorial plane reduces to
	$\Sigma_{\rm eq}=r(r+2b)$. Therefore
	$\Sigma_{\rm eq}=0$ gives $r=0$ and $r=-2b$. Since $b\leq0$, the second
	root lies at a positive radial coordinate for $b<0$. Hence, in the extremal
	case, regularity of the exterior region requires the extremal horizon to lie
	outside this positive equatorial zero, namely $r_E>-2b$. Using
	$r_E=1+b$ for $M=1$, this condition gives
	$1+b>-2b$, and therefore $b>-1/3$. Accordingly, the extremal
	energy-extraction results presented in Table~(\ref{irr_tab}) are restricted to
	the regular exterior domain $-1/3<b\leq0$, which includes all values of
	$b$ considered in the present analysis.
	
	\vspace{0.2cm}
	Consequently, the rotationally extractable energy defined by $\mathcal{E}_{\rm rot}=M-M_{\rm irr}$ increases with decreasing $b$. Numerically, one recovers the standard Kerr extremal value at $b=0$,
	$\mathcal{E}_{\rm rot}\simeq 0.2929$ ($\sim 29\%$)~\citep{chandrasekhar1983mathematical}, whereas for EMDA the extractable fraction grows to $\sim 40\%$ at $b=-0.1$, $\sim 51\%$ at $b=-0.2$, and $\sim 62.6\%$ at $b=-0.3$ (Table~\ref{irr_tab}). 
	Thus, negative dilaton hair progressively lowers the 'irreducible' part of the mass and enhances the maximum reversible rotational-energy reservoir beyond the Kerr bound.
	
	\subsection{Kinematic Bounds on the Local Speeds of the Fragments}\label{sec_4_2}
	
	We have observed so far that, for energy extraction via the Penrose process to occur, the condition $L<0$ must be satisfied. The Wald inequality \citep{1974ApJ...191..231W} and the Bardeen-Press-Teukolsky \citep{1972ApJ178347B} inequality (see \citep{chandrasekhar1983mathematical}) further impose lower bounds on the speed of the ejected particles, ensuring the physical viability of the process. These inequalities serve as a fundamental theoretical tool in analysing the limits of energy extraction from rotating black holes via the Penrose process. In scenarios where a particle disintegrates within the ergoregion, its resulting fragments may carry different energy values. For energy extraction to occur, at least one of these fragments must escape to infinity with a greater total energy than that of the original particle \citep{1986ApJ...307...38P}. The Wald inequality establishes an upper bound on the energy that such fragments can attain, expressed in terms of the underlying spacetime geometry. This constraint plays a pivotal role in quantifying the efficiency of energy extraction processes in rotating spacetimes and helps to elucidate both the origin and limitations of the Penrose process. Whereas the Bardeen-Press-Teukolsky (BPT) inequality provides a kinematic lower bound on the local (relative) speed required for one of the particles produced in the collision to acquire negative Killing energy. In this context, we examine these inequalities within the EMDA spacetime, aiming to demonstrate that the variation in the parameter $b$ interferes with these lower bounds. The derivations presented here closely follow those for the Kerr geometry; therefore, we omit detailed steps and refer to the cited references for a comprehensive treatment.
	
	\subsubsection*{$\bullet$ The Wald Inequality}
	
	Consider a particle (test) moving along a timelike geodesic with four-velocity $u^{\alpha}$, satisfying the normalisation condition $u_{\alpha}u^{\alpha}=-1$, and carrying a conserved energy $\chi>0$. Suppose this particle decays and splits into two fragments. Let $\sigma$ and $v^{\alpha}$, with $v_{\alpha}v^{\alpha}=-1$, denote the conserved energy and four-velocity, respectively, of one of the resulting particles. To analyse the process locally, we introduce an orthonormal tetrad $e_{(\mu)}{}^{\alpha}$, where indices in parentheses label the local Lorentz frame, and choose the timelike basis vector such that $u^{\alpha}=e_{(0)}{}^{\alpha}$~\citep{Pradhan_2019,chandrasekhar1983mathematical,Bhattacharya:2017scw,Rudra:2019ssz}. Expressing the four-velocity $v^{\alpha}$ in the basis of  orthonormal frame $e_{(\mu)}{}^{\alpha}$ as $v^{\alpha} = e_{(\mu)}{}^{\alpha}\,u^{(\mu)}$, we obtain
	
	\begin{equation}
		v^{\alpha} = \frac{ \left(u^{\alpha}+v^{(j)}e_{(j)}{}^{\alpha}\right)}{\sqrt{1- v^{(j)}v_{(j)} }}, \qquad  (j=1,2,3).
		\label{sup22}
	\end{equation}
	
	\vspace{0.2cm}
	Now expanding the timelike killing vector $(\partial_t)^{\alpha}$ in the basis of orthonormal frame, we get
	
	\begin{equation}
		(\partial_t)^{\alpha}=(\partial_t)^{(0)}u^{\alpha}+e_{(j)}{}^{\alpha}(\partial_t)^{(j)}.
		\label{sup23}
	\end{equation}
	
	\vspace{0.2cm}
	Hence, the initial particle's conserved energy $\chi = -\,g_{\alpha \beta}\,u^{\alpha}(\partial_{t})^{\beta}$, can be written in terms of the orthonormal basis as
	
	\begin{equation}
		\chi= (\partial_t)^{(0)}.
		\label{sup24}
	\end{equation}
	
	also
	
	\begin{equation}
		g_{tt}=g_{\alpha \beta}(\partial_t)^{\alpha}(\partial_t)^{\beta}=(\partial_t)^{(j)}(\partial_t)_{(j)}-((\partial_t)^{(0)})^2=(\partial_t)^{(j)}(\partial_t)_{(j)}-\chi^2.
		\label{sup25}
	\end{equation}
	
	\vspace{0.2cm}
	In a similar manner, the conserved energy ($\sigma$) associated with one of the fragments may be expressed as
	
	\begin{equation}
		\sigma=-(\partial_t)^{\alpha}v_{\alpha}= \frac{\chi-{\,|\partial_t| \bf| v|}\cos \varphi }{ \sqrt{1- {\bf |v|}^2} }.
		\label{sup26}
	\end{equation}
	
	\vspace{0.2cm}
	In the orthonormal frame, $\lvert \mathbf{v} \rvert$ denotes the magnitude of the spatial part of the four-velocity, while $\lvert \boldsymbol{\partial}_{t} \rvert$ represents the magnitude of the spatial projection of the timelike Killing vector field; the angle between these two vectors is denoted by $\varphi$. By substituting Eq.~\eqref{sup25} and using the metric component ($g_{tt}=-(\Delta-a^{2}\sin^{2}\theta)/\Sigma$) from Eq.~\eqref{metric}, the preceding relation can be recast as
	
	\begin{equation}
		\sigma= \frac{\chi-{\bf| v|} \left(\chi^2-\frac{\Delta-a^2\sin^2\theta}{\Sigma}\right)^{1/2}\cos \varphi}{\sqrt{1- 
				{\bf |v|}^2 } }.
		\label{sup27}
	\end{equation}
	
	\vspace{0.2cm}
	This leads to the Wald inequality~\citep{1974ApJ...191..231W} in the EMDA spacetime
	
	\begin{equation}
		\frac{\chi-{\bf| v|} \left(\chi^2-\frac{\Delta-a^2\sin^2\theta}{\Sigma}\right)^{1/2}   }{ \sqrt{1- {\bf |v|}^2} }\leq \sigma \leq \frac{\chi+{\bf| v|} \left(\chi^2-\frac{\Delta-a^2\sin^2\theta}{\Sigma}\right)^{1/2}   }{ \sqrt{1- {\bf |v|}^2 } }. 
		\label{sup28}
	\end{equation}
	
	\vspace{0.2cm}
	Thus, in order to have a negative value of $\sigma$, we must have 
	
	\begin{equation}
		{\bf| v|}> \frac{\chi}{\left(\chi^2-\frac{\Delta-a^2\sin^2\theta}{\Sigma}\right)^{1/2}}.
		\label{sup29}
	\end{equation}
	
	\vspace{0.2cm}
	It is evident that the lower bound discussed above attains its minimum at
	$\theta = \pi/2$ and at the black hole horizon, given by
	
	\begin{equation}
		{\bf| v|}> \frac{\chi}{\left(\chi^2+\frac{a^2}{r_{H}^2+2br_{H}}\right)^{1/2}}.
		\label{sup30}
	\end{equation}
	
	\vspace{0.2cm}
	\noindent
	Before the initiation of any energy-extraction process, the fragments must attain relativistic energies, as is evident from Table~(\ref{wald_tab}). The analysis has been further extended to include the extremal configurations, thereby ensuring a more comprehensive treatment. Thus, upon adopting the conventional normalisation $\chi=1$, the maximum energy that a particle can sustain while remaining on a circular orbit around a maximally rotating Kerr black hole is $E_{\max}=1/\sqrt{3}$, as reported in
	\citep{chandrasekhar1983mathematical}. This constraint, in turn, yields a minimum velocity of $|v| = \frac{1}{\sqrt{2}}$, which is necessary to effectively initiate the Penrose process, as originally derived by Chandrasekhar \citep{chandrasekhar1983mathematical}. It is observed that EMDA black hole achieves systematically lower minimum speeds as $b$ decreases toward $-1/3$, consistent with the same trend that yields much higher maximal extraction efficiencies in the EMDA family near that limit.
	\begin{table}[h!]
		\centering
		\begin{tabular}{|c|c|c|c|}
			\hline
			$\bf b$ & $\bf a_E$ & $\bf r_{E}$ & $\bf \mathbf{|v|_{\min}}$ \\ \hline \hline
			0.0  & 1.000000 & 1.000000 & 0.70710678 \\
			-0.1  & 0.900000 & 0.900000 & 0.66143783 \\
			-0.2  & 0.800000 & 0.800000 & 0.57735027 \\
			-0.3  & 0.700000 & 0.700000 & 0.35355339 \\ \hline
		\end{tabular} 
		\caption{Lower bounds on the local speed $\lvert v\rvert_{\min}$ of the fragments for different values of the dilaton parameter $b$ and the spin parameter $a$ in the extremal limit.} \label{wald_tab}
	\end{table}
	
	\subsubsection*{$\bullet$ The Bardeen-Press-Teukolsky Inequality}
	
	The Bardeen-Press-Teukolsky (BPT) inequality~\citep{1972ApJ178347B,Pradhan_2019,Bhattacharya:2017scw,chandrasekhar1983mathematical} yields results analogous to those discussed above. Consider two particles with energies $\sigma_{+}$ and $\sigma_{-}$ that collide at a specific spacetime point. An orthonormal basis $e_{(\mu)}{}^a$ is constructed as before, with the timelike vector $e_{(0)} = u^a$ representing the four-velocity of the local observer. In this locally inertial frame, we assume that the particles move with equal and opposite spatial velocities, denoted by $v^{(i)}$ and $-v^{(i)}$. The magnitude of their local relative speed in this frame can be determined using the (special) relativistic velocity addition formula. Given that the particles move with equal and opposite three-velocities, the relative speed $| v_{\rm rel}|$ between them is given by
	
	\begin{equation}
		{\bf| v_{\rm rel}|}= \frac{2 {\bf |v|}}{1+{\bf |v|}^2}.
		\label{sup31}
	\end{equation}
	
	\vspace{0.2cm}
	\noindent
	Here, $\mathbf{|v|}$ denotes the magnitude of the three-velocity of each particle, as measured in the local orthonormal frame. Our objective is to determine a lower bound on the relative velocity $|\mathbf{v}_{\rm rel}|$ such that the energy $\sigma_{-}$ of one of the colliding particles becomes negative. By following a procedure similar to that outlined in the previous subsection, one arrives at the following result.
	
	\begin{equation}
		\sigma_{\pm}=\frac{(\partial_t)^{(0)}\pm {\,|\partial_t| \bf| v|}\cos \varphi   }{ \sqrt{1- {\bf |v|}^2} }.
		\label{sup32}
	\end{equation}
	
	\vspace{0.2cm}
	Using the above expression, and after some algebraic manipulation, we obtain the following result, 
	
	\begin{equation}
		\left(\sigma_{+}+\sigma_{-}\right)^2\:{\bf |v|}^4-2\left(\sigma_{+}^2++2g_{tt}+\sigma_{-}^2\right){\bf |v|}^2+(\sigma_{+}-\sigma_{-})^2\leq 0,
		\label{sup33}
	\end{equation}
	which gives
	\begin{equation}
		{\bf |v|}\geq \frac{\sqrt{\sigma_{+}^2-\left(\frac{\Delta-a^2\sin^2\theta}{\Sigma}\right)}-\sqrt{\sigma_{-}^2-\left(\frac{\Delta-a^2\sin^2\theta}{\Sigma}\right)} }{\left(\sigma_{+}+\sigma_{-}\right)}.
		\label{sup34}
	\end{equation}
	
	\vspace{0.2cm}
	Considering the marginal case where $\sigma_{-} = 0$, we evaluate the condition on the horizon at $\theta = \pi/2$,
	
	\begin{equation}
		{\bf |v|}\geq \frac{1}{\sigma_+}\left(\sqrt{\sigma_+^2+\frac{a^2}{r_{H}^2+2br_{H}}}-\sqrt{\frac{a^2}{r_{H}^2+2br_{H}}} \right).
		\label{sup35}
	\end{equation}
	
	\vspace{0.2cm}
	Partially differentiating this with respect to $r_H$, we find
	
	\begin{equation}
		\frac{\partial |v|_{\min}}{\partial r_{H}}
		\geq \frac{a^{2}(r_{H}+b)}{(r_{H}^{2}+2br_{H})^{2}}
		\left(
		\sqrt{\frac{r_{H}^{2}+2br_{H}}{a^{2}}}
		-\frac{1}{\sqrt{1+\dfrac{a^{2}}{\,r_{H}^{2}+2br_{H}\,}}}
		\right) \geq 0.
	\end{equation}
	
	\begin{table}[h!]
		\centering 
		\begin{tabular}{|c|c|c|c|c|}
			\hline
			$\bf b$ & $\bf a_E$ & $\bf r_E$ & $\bf \mathbf{|v|_{\min}}$ & $\bf \mathbf{|v|_{\text{rel}}}$ \\
			\hline \hline
			0.0  & 1.0000000 & 1.0000000 & 0.4142136 & 0.7071068 \\
			-0.1 & 0.9000000 & 0.9000000 & 0.3779645 & 0.6614378 \\
			-0.2 & 0.8000000 & 0.8000000 & 0.3178372 & 0.5773502 \\
			-0.3 & 0.7000000 & 0.7000000 & 0.1826758 & 0.3535533 \\
			\hline
		\end{tabular}
		\caption{Minimum values of the local fragment speed ($\lvert v\rvert_{\min}$) and the relative speed $\lvert v\rvert_{\text{rel}}$ corresponding to different choices of the parameters $b$ and $a$ in the extremal configuration.}
		\label{bar_tab}
	\end{table}
	
	\vspace{0.1cm}
	\noindent
	Finally, we once again adopt the standard choice $\sigma_{+}=1$. For the extremal Kerr black hole ($b=0$), we obtain $\lvert v\rvert_{\min}\ge 0.414$, which, using Eq.~(\ref{sup31}), implies $\lvert v_{\rm rel}\rvert_{\min}\ge 0.707$ (see also \citep{chandrasekhar1983mathematical}). For the extremal EMDA solution, the threshold decreases as the dilaton parameter becomes more negative; for instance, for $b=-0.3$ we find $\lvert v\rvert_{\min}\ge 0.183$, leading to $\lvert v_{\rm rel}\rvert_{\min}\ge 0.354$ (see Table~\ref{bar_tab}). Thus, a negative dilaton parameter $b$ significantly lowers the kinematic threshold relevant for energy-extraction processes compared to the Kerr case, and this reduction becomes particularly pronounced when the spin approaches its allowed maximum for a given $b$.
	
	\vspace{0.3cm}
	\noindent
	We next investigate the physical origin of the reduction in the kinematic thresholds as the dilaton parameter $b$ assumes increasingly negative values. From the Wald inequality in Eq.~\eqref{sup30}, the lower bound on the local fragment speed at extremality takes the closed form $|v|_{\min}^{\rm Wald}=\sqrt{\frac{1+3b}{2(1+2b)}}$, while the BPT inequality in Eq.~\eqref{sup35} yields $|v|_{\min}^{\rm BPT}=\frac{\sqrt{2(1+2b)}-\sqrt{1+b}}{\sqrt{1+3b}}$. These reductions are not primarily due to a widening of the ergoregion. Indeed, at extremality the equatorial ergosurface and the horizon are located at $r_{\rm ergo}=2(1+b),\: r_E=1+b$, so that $r_{\rm ergo}/r_E=2$, independent of $b$. Moreover, the coordinate width $r_{\rm ergo}-r_E=1+b$ decreases as $b$ becomes more negative. The dominant contribution therefore arises from the near-horizon rotational geometry, encoded in the horizon angular velocity $\Omega_E=\frac{1}{2(1+2b)}$. Within the admissible regular exterior domain, $-1/3<b\leq0$, $\Omega_E$ increases as $b$ decreases. Since the extremal spin is $a_E=1+b$, the Wald threshold may equivalently be written as $|v|_{\min}^{\rm Wald}=\sqrt{1-a_E\Omega_E}$. Thus, as $b$ becomes more negative, the near-horizon rotational factor $a_E\Omega_E$ increases, the quantity $1-a_E\Omega_E$ decreases, and the minimum local velocity required to produce a negative-energy fragment is correspondingly lowered. Hence the enhanced kinematic permissiveness of the EMDA spacetime relative to Kerr is driven primarily by the strengthened near-horizon frame-dragging contribution, rather than by a simple change in the ergoregion geometry.
	
	\section{Superradiance of EMDA Black Hole}\label{sec_5}
	
	The Penrose mechanism operates under the condition that the fragment escaping from a rotating black hole possesses a conserved energy greater than that of the initial particle, as described in the previous section. However, because the required astrophysical conditions are difficult to realise, the Penrose process is expected to be inefficient in realistic astrophysical environments. Superradiance, by contrast, is a wave phenomenon: although dissipation can enhance it in practical settings by providing an effective amplifier, superradiance can occur even in a vacuum provided the background spacetime is curved and admits the appropriate scattering conditions. In this sense, superradiance~\citep{pengpan2025superradiancequasinormalmodesmassive,Mollicone_2025,Liu_2024,Jha_2024,Mukherjee_2019,Laurent_Di_Menza_2015,Cheriyodathillathu_2025,herrerovalea2024superradiantscatteringlorentzviolatinggravity,Brito_2020} may be viewed as the wave analogue of the Penrose process. When an incident wave scatters off a rotating black hole, it is partially reflected and partially absorbed. Under superradiant conditions, the reflected wave emerges with enhanced amplitude, whereas the absorbed portion carries negative-energy as measured by an observer at infinity.
	
	\vspace{0.2cm} 
	\noindent
	In this section, we investigate the superradiant scattering of radiation in the EMDA black hole spacetime. For this purpose, we consider a neutral, massless, minimally coupled external test scalar field $\Phi$ propagating on a fixed EMDA black hole background. It is noteworthy to emphasize that this probe field is distinct from the dilaton field ($\phi$) and the axion field ($\kappa$) that appear in the EMDA action. The dilaton and axion are intrinsic components of the background solution, and their perturbations are generally coupled to the electromagnetic and metric perturbations through the full EMDA field equations. Consequently, a complete analysis of the physical dilaton-axion perturbation sector would require solving a coupled system involving gravitational, electromagnetic, dilatonic, and axionic degrees of freedom. In the present work however, we restrict our attention to the simpler probe-field model. Hence it allows us to isolate the role of the EMDA geometry, particularly the dilaton parameter $b$, in modifying the standard superradiant condition. Furthermore, the scalar field $\Phi$ is assumed to carry negligible stress-energy compared with the black hole background, so that its back-reaction on the spacetime geometry can be consistently neglected. 
	
	\vspace{0.2cm}
	\noindent
	Following the heuristic approach described in~\citep{Padmanabhan:2010zzb}, we start by introducing the continuity equation associated with the conserved current.
	
	\begin{equation}
		\Box \Phi= \frac{1}{\sqrt{-g}} \partial_\mu ( \sqrt{-g}g^{\mu\nu} \partial_\nu \Phi)=0,
	\end{equation} 
	
	\vspace{0.2cm}
	we adopt the usual separable mode ansatz
	
	\begin{equation}
		\Phi=e^{-i\beta t}e^{ik\phi}\Theta(\theta)R(r),
	\end{equation} 
	
	\vspace{0.2cm}
	where $\beta$ is the wave frequency and $k$ is the azimuthal quantum number.
	
	\vspace{0.2cm}
	\noindent
	Following the standard procedure mentioned in~\citep{Padmanabhan:2010zzb}, the rate of energy loss, or equivalently the associated power flux, takes the form
	
	\begin{equation}
		dP=\beta(\beta-k \Omega_H) \left(\frac{r^2+2br+a^2}{\Sigma}\right)_{r_E} \iint (\Sigma)_{r_E} \Theta(\theta)^2 sin^2 \theta d\theta d\phi,
	\end{equation}
	
	or,
	\begin{equation}
		\begin{aligned}
			P &= \beta(\beta-k\Omega_H)[r_{E}^2+2br_{E}+a^2], \\[10pt]
			&= constant. \label{13}
		\end{aligned}
	\end{equation}
	
	\vspace{0.2cm}
	Here $\Omega_H=\frac{a}{\Xi}$ is the angular velocity of the outer horizon. The superradiance is impossible if $\beta>k \Omega_H$, when $P$ is positive. However, on the other hand, the superradiance will occur if $\beta$ lies in the range $0<\beta<k \Omega_H$. Within this range, Eq.~\eqref{13} gives a negative horizon flux, indicating superradiant energy extraction from the black hole. The angular momentum quantum number ($k$) must be non-zero, as it has to remove angular momentum from the black hole. Hence the energy flux lost per unit time is $P(\beta) = \Xi \beta(\beta - k\Omega_H)$, which attains its minimum at $\beta_{\min}=\frac{k a}{2\Xi}$ with value $P_{\min}=\frac{-(k a)^2}{(4\Xi_H)}$.
	
	\begin{figure}[h!]
		\begin{centering}
			\includegraphics[width=8cm,height=7.8cm]{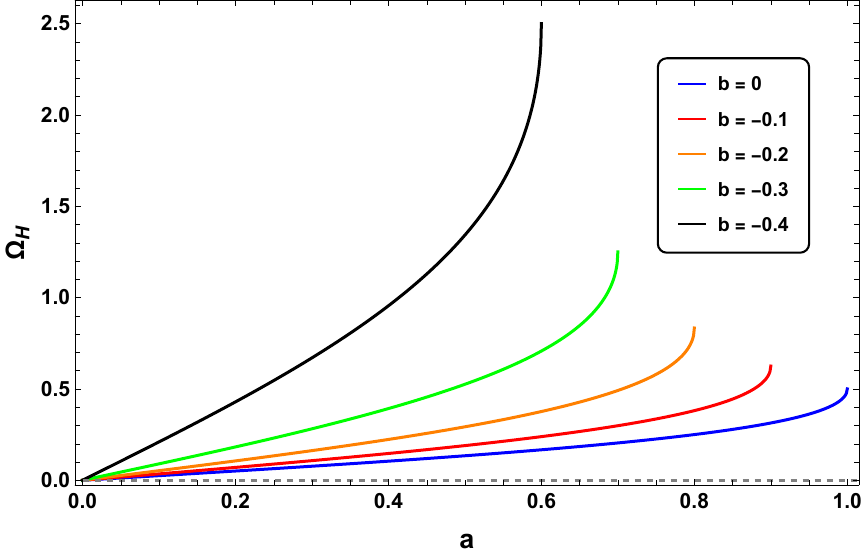}
			\par\end{centering}
		\caption{Horizon angular velocity $\Omega_H$ versus the spin parameter $a$ for selected values of the dilaton parameter $b$ (with $M=1$). Negative $b$ increases $\Omega_H$ for fixed $a$ by reducing the horizon radius $r_H$ and the metric component $\Xi$. The curves terminate at the extremal values $a_E = 1+b$ for each $b$.} 
		\label{velo_plot}
	\end{figure}
	
	\begin{table}[ht!]
		\centering
		\renewcommand{\arraystretch}{1.2}
		\begin{tabular}{|c|c|c|c|c|c|}
			\hline
			\bf Case 
			& $\bf \boldsymbol{\Xi_H}$ 
			& $\bf \boldsymbol{\Omega_H}$ 
			& $\bf \boldsymbol{\beta_{\min}=k\Omega_H/2}$ 
			& $\bf \boldsymbol{\beta_{cut}=k\Omega_H}$ 
			& $\bf P_{\min}$ \\
			\hline \hline
			A ($a=0.6,\,b=-0.2$) & 1.594980 & 0.376180 & 0.376180 & 0.752360 & $-0.225708$ \\
			B ($a=0.4,\,b=-0.4$) & 0.418885 & 0.954915 & 0.954915 & 1.909830 & $-0.381966$ \\
			C ($a=0.8,\,b=-0.1$) & 2.099697 & 0.381007 & 0.381007 & 0.762015 & $-0.304806$ \\
			D ($a=0.7,\,b=-0.3$) & 0.560000 & 1.250000 & 1.250000 & 2.500000 & $-0.875000$ \\
			\hline
		\end{tabular}
		\caption{Superradiant flux characteristics for four representative EMDA black hole configurations. The numerical values are computed for $M=1$ and $k=2$. For each case, the outer horizon is obtained from $r_H=(1+b)+\sqrt{(1+b)^2-a^2}$, and the horizon angular velocity is $\Omega_H=\frac{a}{\Xi_H}, \Xi_H=r_H^2+2br_H+a^2$. The flux minimum occurs at $\beta_{\min}=\frac{k\Omega_H}{2}$, while the upper edge of the superradiant window is $\beta_{cut}=k\Omega_H$. The minimum flux is $P_{\min}=-\frac{(ka)^2}{4\Xi_H}.$ Negative $P_{\min}$ indicates negative horizon flux and hence energy extraction from the black hole. As $b$ becomes more negative and the configuration approaches extremality, $\Omega_H$ increases, the superradiant window broadens, and the negative horizon-flux minimum becomes deeper.}\label{sup_tab}
	\end{table}
	
	\begin{figure}[tbp!]
		\centering
		\subfigure[]{\includegraphics[width=7.6cm,height=7.8cm]{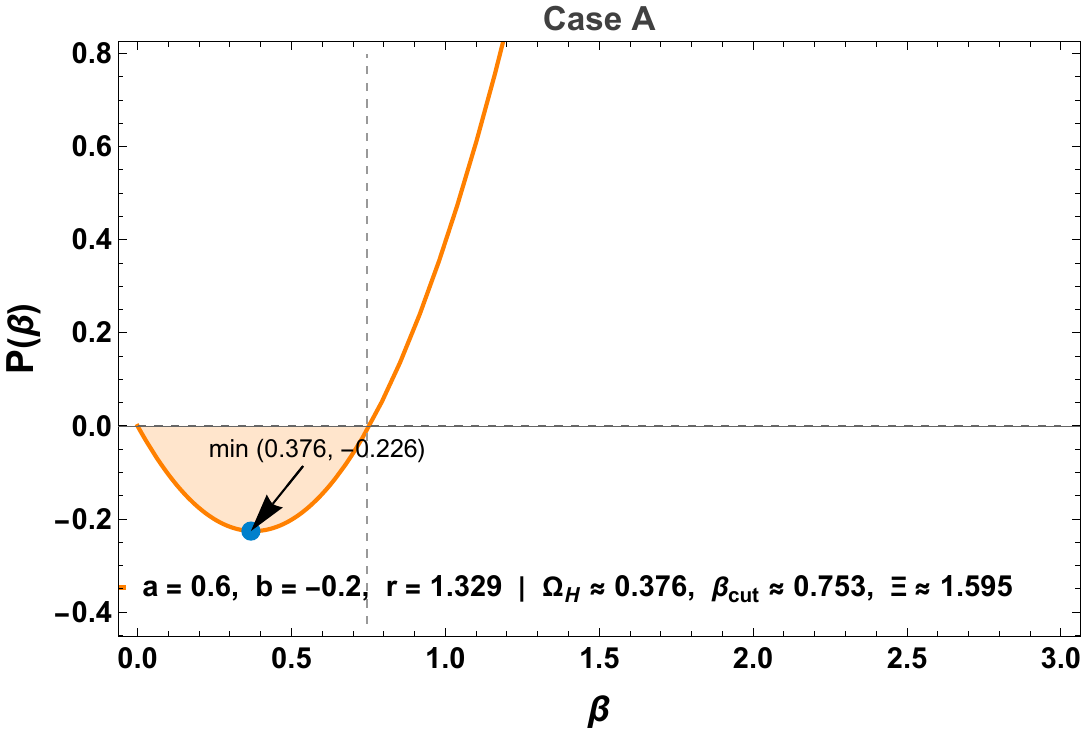}\label{sup_a}}
		\hspace{0.8cm}
		\subfigure[]{\includegraphics[width=7.6cm,height=7.8cm]{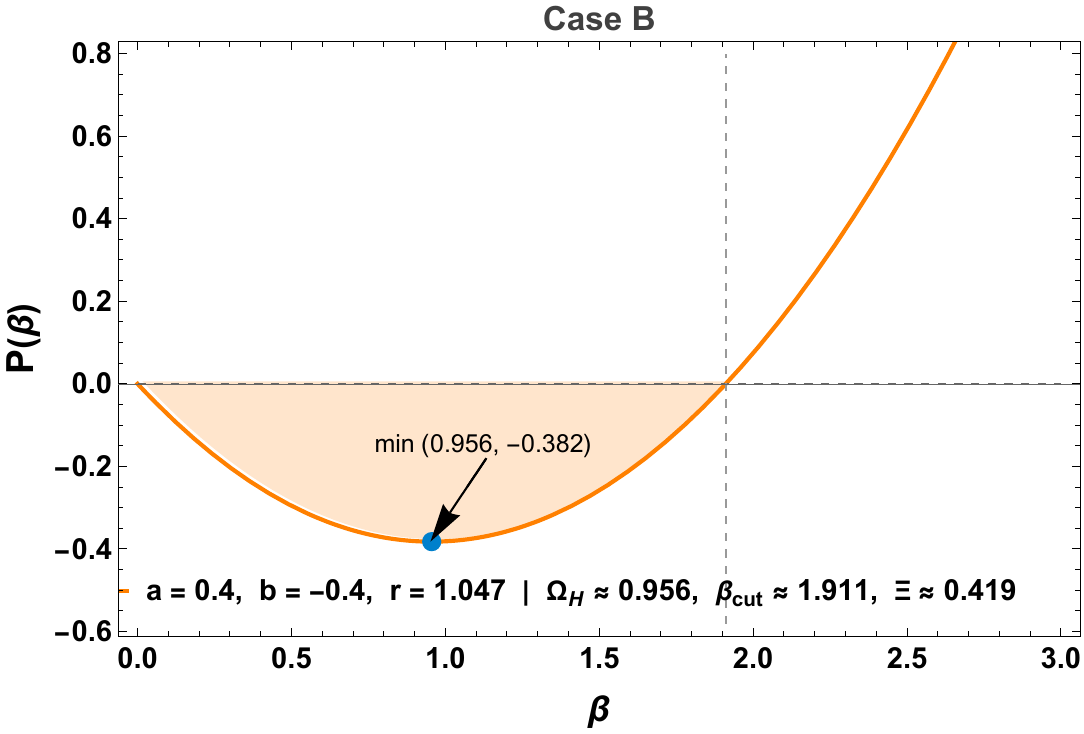}\label{sup_b}}
		\vspace{0.8em} 
		\subfigure[]{\includegraphics[width=7.6cm,height=7.8cm]{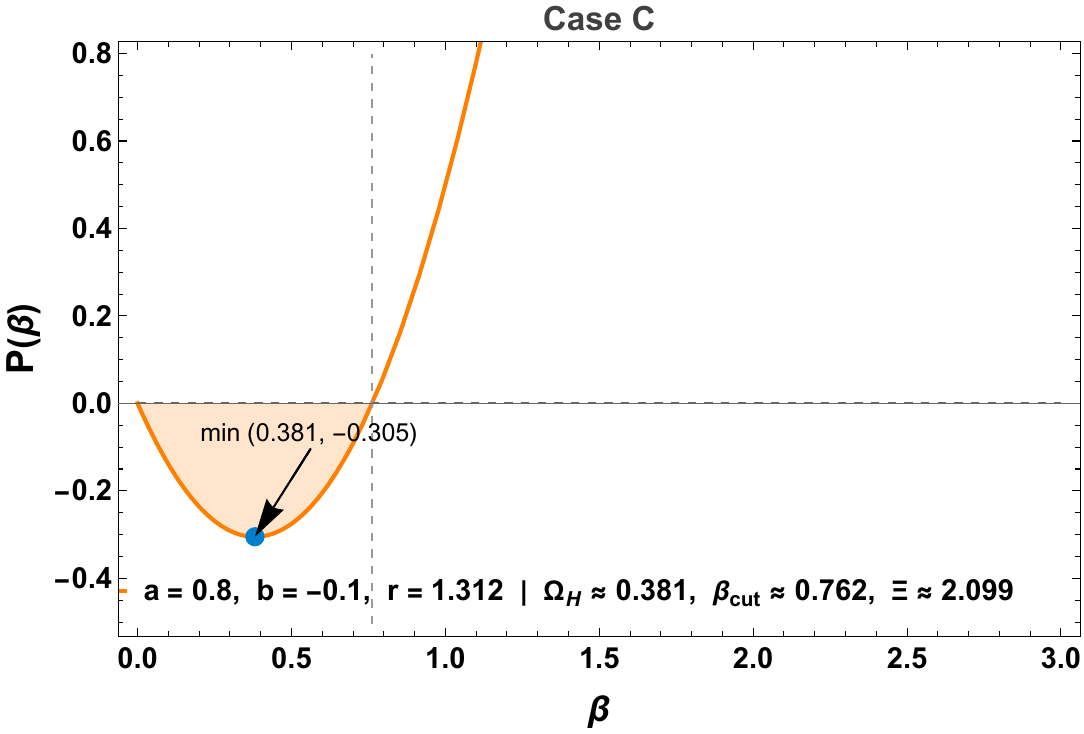}\label{sup_c}}
		\hspace{0.8cm}
		\subfigure[]{\includegraphics[width=7.6cm,height=7.8cm]{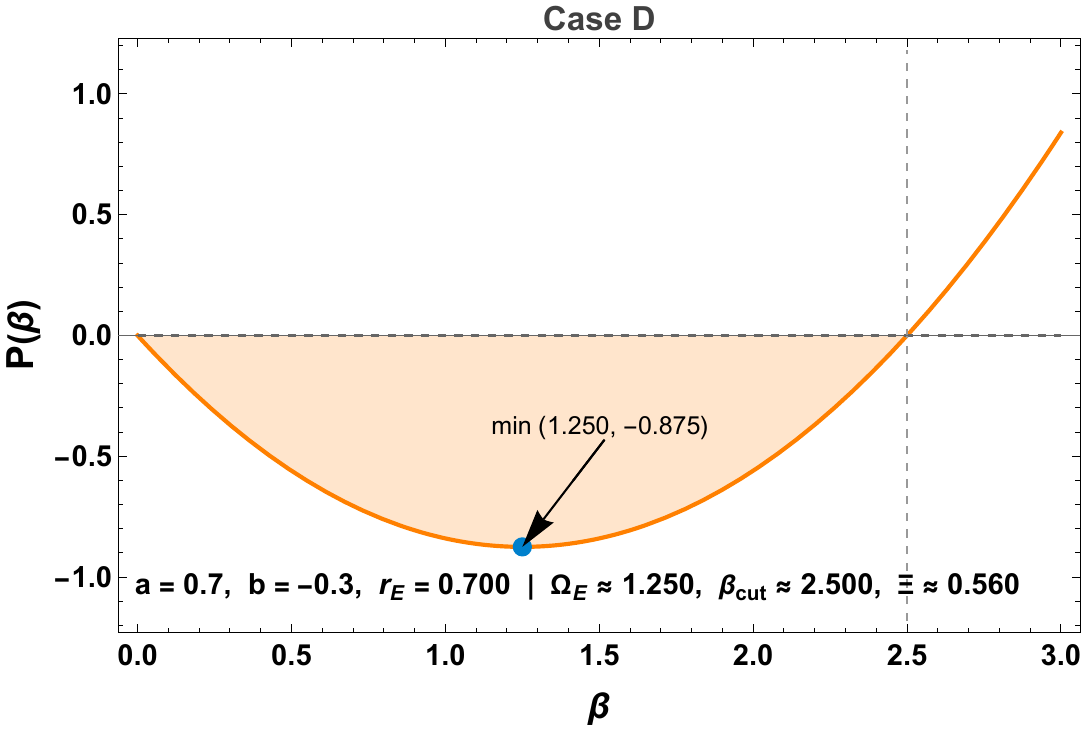}\label{sup_d}}
		\caption{Superradiant horizon flux $P(\beta)$ versus wave frequency $\beta$ for four representative EMDA black hole configurations. The shaded region in each panel denotes the superradiant negative-flux interval $0<\beta<(\beta_{cut}=k\Omega_H$), where $P(\beta)<0$. The boundary $\beta_{\rm cut}=k\Omega_H$ marks the upper edge of the superradiant window, while the minimum occurs at $\beta_{\min}=k\Omega_H/2$. Here, $\Omega_H=a/\Xi_{H}$ is the angular velocity of the outer event horizon, with $\Xi_H$ denoting the horizon value of the metric component $\Xi$. The four panels correspond to:
				(A) $a=0.6$, $b=-0.2$, $r_H=1.329$;
				(B) $a=0.4$, $b=-0.4$, $r_H=1.047$;
				(C) $a=0.8$, $b=-0.1$, $r_H=1.312$; and
				(D) at extremality with $a_E=0.7$, $b=-0.3$, $r_E=0.7$. The numerical values of $\Omega_H$,
				$\beta_{\min}$, $\beta_{\rm cut}$, and $P_{\min}$ are listed in Table~(\ref{sup_tab}).}\label{sup}
	\end{figure}
	
	\vspace{0.2cm}
	A detailed comparison of the four configurations (Cases A, B, C and D) is illustrated in Fig.~(\ref{sup}). The shaded region in each panel denotes the superradiant negative-flux domain, $0<\beta<\beta_{\rm cut}=k\Omega_H$, where the horizon flux satisfies $P(\beta)<0$. The unshaded region corresponds to $P(\beta)>0$, for which superradiant extraction does not occur. The analysis of the superradiant flux reveals a clear and systematic trend (see Table~\ref{sup_tab}). The strength of the negative horizon-flux regime is governed jointly by the spin parameter $a$, the dilaton charge $b$, and the horizon factor $\Xi_H=r_H^2+2br_H+a^2$. In Cases A and C (see Figs.~\ref{sup_a} and \ref{sup_c}), where $b$ is only weakly negative, the horizon radius remains relatively large. This gives moderate values of $\Omega_H$ and consequently narrower superradiant windows, with $k\Omega_H\sim0.75$. The corresponding flux minima are relatively shallow, $P_{\min}\approx -0.23\rightarrow -0.30$, indicating a weaker negative horizon-flux regime. By contrast, Case B (Fig.~\ref{sup_b}) shows that increasing the negativity of $b$ to $-0.4$ reduces $\Xi_H$, thereby increasing the horizon angular velocity to $\Omega_H\approx0.95$ and broadening the superradiant window. The associated minimum, $P_{\min}\approx-0.38$, is deeper, showing that the dilaton parameter can enlarge the negative horizon-flux regime even for a lower spin. This effect is most pronounced in Case D (Fig.~\ref{sup_d}), corresponding to the extremal configuration $(b=-0.3,a_E=0.7)$, where the horizon radius shrinks to $r_E=0.7$. The resulting decrease in $\Xi_H$ gives a large horizon angular velocity, $\Omega_E=1.25$, the widest superradiant interval,
	$k\Omega_E=2.5$, and the deepest flux minimum, $P_{\min}=-0.875$.
	
	\vspace{0.2cm}
	\noindent
	The enhancement discussed here should be understood as an enlargement of the superradiant frequency window and of the negative horizon-flux regime, rather than as a full computation of the greybody amplification factor. Therefore, negative values of $b$ modify the superradiant interval mainly by changing the horizon radius and reducing the horizon factor $\Xi_H$, which in turn increases $\Omega_H$. In this sense, the widening of the superradiant window is primarily driven by the increase of the horizon angular velocity. The scalar-wave effective potential is also modified by the EMDA metric functions and may affect the detailed transmission and reflection coefficients. However, in the present work we do not solve the full radial scattering problem with asymptotic boundary conditions, and therefore we do not compute an explicit greybody factor. The present analysis infers superradiance from the sign of the horizon flux. 
	Taken together, these results indicate that, for the representative EMDA
	configurations considered here, a negative dilaton parameter and proximity to
	extremality can increase the horizon angular velocity, thereby widening the
	neutral-scalar superradiant frequency window and deepening the corresponding
	negative horizon-flux minimum relative to the Kerr limit~\citep{Padmanabhan:2010zzb}. In other words, while spin alone increases efficiency, it is the combined interplay of $a$, $b$, and the resulting horizon structure that controls the overall strength of superradiance in EMDA spacetimes, with the extremal limit representing the most favourable scenario for energy extraction.
	
	\section{Center-of-Mass Energy of Two Colliding Particles in the EMDA Black Hole}\label{sec_6}
	
	In a seminal study, Bañados, Silk, and West \citep{PhysRevLett.103.111102} examined the collision of two particles in the vicinity of a Kerr black hole and proposed that a rotating black hole could effectively act as a natural particle accelerator. Their analysis revealed that the center-of-mass energy $(E_{CM})$ of two colliding particles in the equatorial plane can, in principle, become unbounded in the extremal limit of the black hole. This remarkable result implies that an extremal Kerr black hole can accelerate particles to energies approaching the Planck scale, opening new avenues for exploring ultra-high-energy astrophysical phenomena such as gamma-ray bursts, active galactic nuclei, and other extreme collision processes. Consequently, the BSW mechanism has garnered substantial attention \citep{Amir_2016,PhysRevD.82.103005,Mukherjee_2019,Liu:2010ja,Mao_2017,Amir:2015pja,Zakria:2015eua,PhysRevD.111.024022,PhysRevD.110.064016,Ovcharenko_2024,Zhang_2025,Rudra:2019ssz,Babar:2021nst,Zaslavskii_2010,Zaslavskii_2023,Patil:2011aab,Patil:2011aaa,Pradhan:2014oaa,Sheoran:2020kmn,PhysRevD.90.103006,Harada_2014} as a fundamental framework for understanding high-energy particle interactions and energy extraction in the strong-gravity regime of rotating black holes. 
	
	\vspace{0.2cm}
	\noindent
	In particular, Hussain~\citep{2012MPLA...2750068H} investigated the collision of two uncharged particles freely falling from rest at infinity along radial trajectories in the background of an extremal EMDA black hole. In that analysis, the azimuthal motion is suppressed, and consequently no independent conserved particle angular momentum $L_i$ appears in the final expression for $E_{CM}$. Under this restricted radial assumption, the center-of-mass energy remains finite at the event horizon. Moreover, when the spin and dilaton parameters are taken at their limiting values, the horizon value of $E_{CM}$ remains finite and does not exhibit any explicit dependence on the black hole parameters. The present work revisits the same problem in a more general equatorial setting, where the colliding particles are allowed to carry conserved angular momenta $L_1$ and $L_2$. In this formulation, the dilaton parameter $b$ enters explicitly into the expression for $E_{CM}$ through the EMDA metric functions, the horizon radius, and the horizon angular velocity. In the noncritical sector, and in particular in the corresponding radial/zero-angular momentum limit, our general expression reduces to a finite horizon value, in agreement with Hussain's result. However, once the angular-momentum degree of freedom is retained, an additional critical channel becomes available. The near-horizon behaviour is governed by $E_i-\Omega_H L_i$, so that the BSW critical angular momentum is $L_c$. We show that $E_{CM}$ remains finite for generic angular momenta $L_i\neq L_c$, but diverges in the extremal near-horizon limit when one of the particles is tuned to $L_i=L_c$. Thus, even after incorporating the explicit $b$-dependence of the EMDA geometry into $E_{CM}$, the BSW divergence persists. The role of $b$ is not to remove the divergence, but to shift the horizon structure, modify $\Omega_H$, and consequently change the critical angular momentum $L_c$. Therefore Hussain's finite result is not contradicted; it is recovered in the radial/noncritical sector, while the present analysis extends the calculation to the critical angular-momentum sector of the EMDA black hole.
	
	\vspace{0.2cm}
	\noindent
	In this section, we examine the behaviour of the $E_{CM}$ of colliding particles in the vicinity of a rotating EMDA black hole, particularly in the limiting case where the radial coordinate $r$ approaches the event horizon $r_{H}$. Let us consider two uncharged particles of equal rest mass $M_{1} = M_{2} = M_{0}$ that are released from rest at infinity, such that $E_{1}/M_{0} = E_{2}/M_{0} = 1$. These particles approach the rotating EMDA black hole with different specific angular momenta $L_{1}$ and $L_{2}$, and collide at some finite radial coordinate $r$. The motion of the two particles can be described by their respective four-momenta, each defined through the conserved energy and angular momentum in the EMDA spacetime as
	
	\begin{equation}
		P_{i}^{\mu} = M_{i} u_{i}^{\mu}, \label{ecm_1}
	\end{equation}
	
	\vspace{0.2cm}
	where $u^\mu_{(i)}$ denotes the physical four-velocity of the particles ($i=1,2$) with respect to the proper time $\alpha$, such that $u^\mu_{(i)}=dx^\mu_{(i)}/d\alpha$. Although the geodesic equations in Section~(\ref{sec_3}) are expressed in terms of the Mino parameter $\tau$, the corresponding physical four-velocities are obtained from the Mino-time velocities through $u^\mu_{(i)}=\Sigma^{-1}dx^\mu_{(i)}/d\tau$. The center-of-mass energy ($E_{\mathrm{cm}}$) of the two colliding particles is then given by
	
	\begin{equation}
		E_{\mathrm{cm}}^{2} = -\,P_{i}^{\mu} P_{i\mu}.\label{ecm_2}
	\end{equation}
	
	\vspace{0.2cm}
	By substituting Eq.~(\ref{ecm_1}) into Eq.~(\ref{ecm_2}), we obtain
	
	\begin{equation}
		\frac{E_{\mathrm{cm}}^{2}}{2M_{0}^{2}} = 1 - g_{\mu\nu}\,u_{(1)}^{\mu}u_{(2)}^{\nu}.\label{ecm_3}
	\end{equation}
	
	\vspace{0.2cm}
	\noindent
	Now, substituting the components of $g_{\mu\nu}$, $u_{(1)}^{\mu}$, and $u_{(2)}^{\mu}$ for the spacetime considered here, the center-of-mass energy simplifies to
	
	\begin{align}
		\frac{E_{\rm CM}^2}{2M_0^2}
		&=\frac{1}{r(2b+r)\Big(a^2+r\big(r-2(1+b)\big)\Big)}
		\Bigg(
		2ra^2(r+4b+1)
		-2ra(1+2b)(L_1+L_2) \nonumber\\[10pt] 
		&\hspace{2.2cm}
		+r(2b-r+2)L_1L_2
		+2r(r-1)r(2b+r)
		-\sqrt{B_1}\,\sqrt{B_2}
		\Bigg), \label{Ecm_EMDA}\\[10pt]
		B_i
		&= r\Big(
		a^2(4b+2)
		-4a(2bL_i+L_i)
		+L_i^2(2b-r+2)
		+2(2b+1)r(2b+r)
		\Big),\qquad i=(1,2).\nonumber
	\end{align}
	
	\vspace{0.2cm}
	Since $E_{CM}$ is an invariant scalar, it represents a physical observable and does not depend on the choice of coordinates. This guarantees the accuracy of the formula given in Eq.~(\ref{Ecm_EMDA}) in both general relativity and special relativity with $M_0 = 1$, $M$ = 1. The result obtained in Eq. (\ref{Ecm_EMDA}) indicates that the parameter $b$ has a significant influence on the $E_{CM}$. Notably, in the limiting case where the dilaton parameter vanishes $b = 0$, the corresponding expression for the $E_{CM}$ reduces to the well-established result obtained for the Kerr black hole, as reported in \citep{PhysRevLett.103.111102}.
	
	\vspace{0.2cm}
	\noindent
	To determine the allowed range of angular momentum for particles colliding near the black hole's horizon, we first compute the corresponding effective potential. The radial equation governing the motion of a timelike particle along a geodesic can be expressed as
	
	\begin{equation}
		\frac{1}{2} {(u^{r})}^{2}+V_{eff}=0.
	\end{equation}
	
	\vspace{0.3cm}
	The effective potential $V_{eff}$ for EMDA black hole at the equatorial plane reads as
	
	\begin{equation}
		V_{eff} = -\frac{\big(\Xi E - aL\big)^2- \Delta \Big[(aE - L)^2 + m_0^{2}\big(r^2 + 2 b r\big)\Big]}
		{2(r^2+2br)^{2}}.
	\end{equation}
	
	\vspace{0.3cm}
	Now, the condition for circular orbits of test particles in the rotating EMDA black hole spacetime is obtained by solving
	
	\begin{equation}
		V_{eff}=0, \hspace{3mm}  \text{and} \hspace{3mm} \frac{dV_{eff}}{dr}=0. \label{l_val}
	\end{equation}
	
	\vspace{0.2cm}
	\begin{figure}[h!]
		\begin{centering}
			\subfigure[]{\includegraphics[width=7.4cm,height=7.8cm]{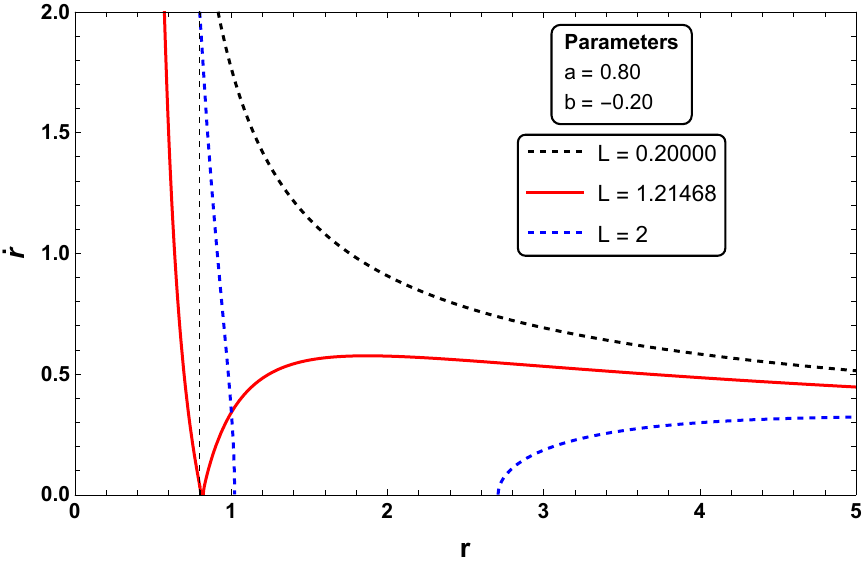}\label{rdot_1}} \hspace{0.8cm}
			\subfigure[]{\includegraphics[width=7.4cm,height=7.8cm]{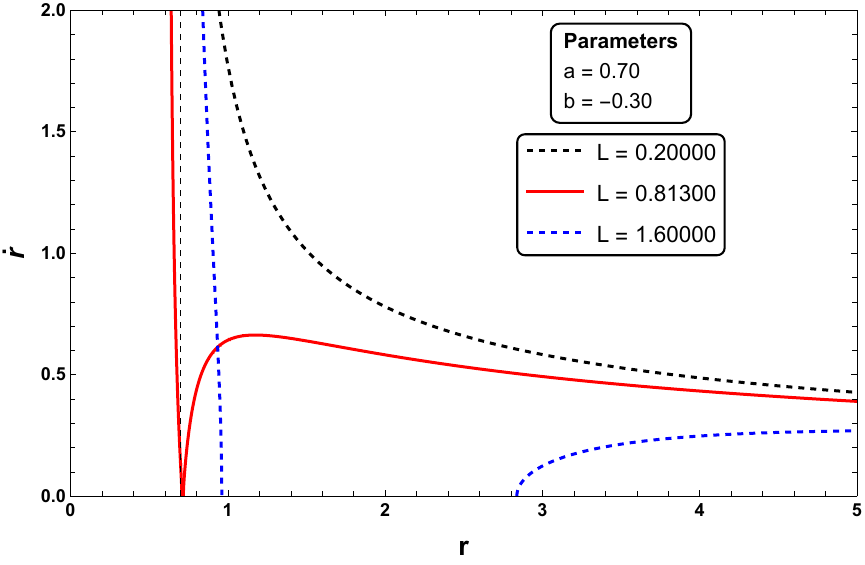}\label{rdot_2}}
			\par\end{centering}
		\caption{Illustration of the radial velocity $\dot{r}$ versus the radial coordinate $r$ for an extremal EMDA black hole for different values of angular momentum $L$, $b$ and $a$. The black dotted lines correspond to $L > L_c$, the solid red line denotes $L = L_c$, and the blue dotted lines indicate $L < L_c$. The vertical dashed line denotes the location of the extremal event horizon $r_E$ of the EMDA black hole.}\label{rdot_plot}
	\end{figure}
	
	\begin{figure}[h!]
		\centering
		\subfigure[]{\includegraphics[width=7.5cm,height=7.8cm]{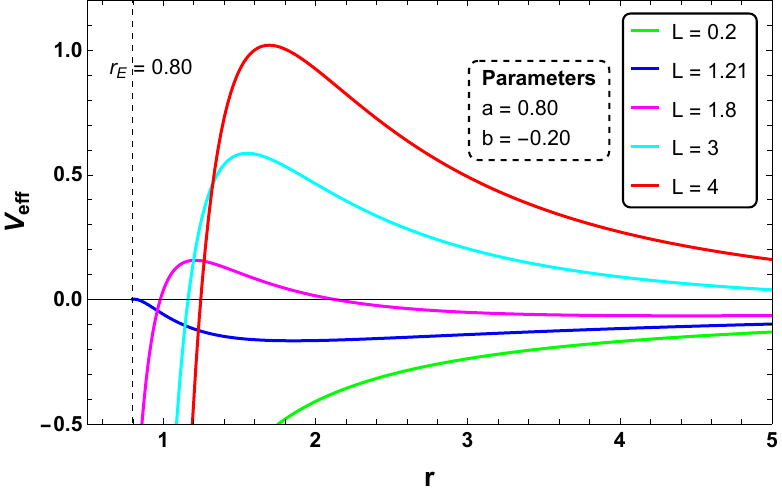}\label{veff_1}}\hspace{0.8cm}
		\subfigure[]{\includegraphics[width=7.5cm,height=7.8cm]{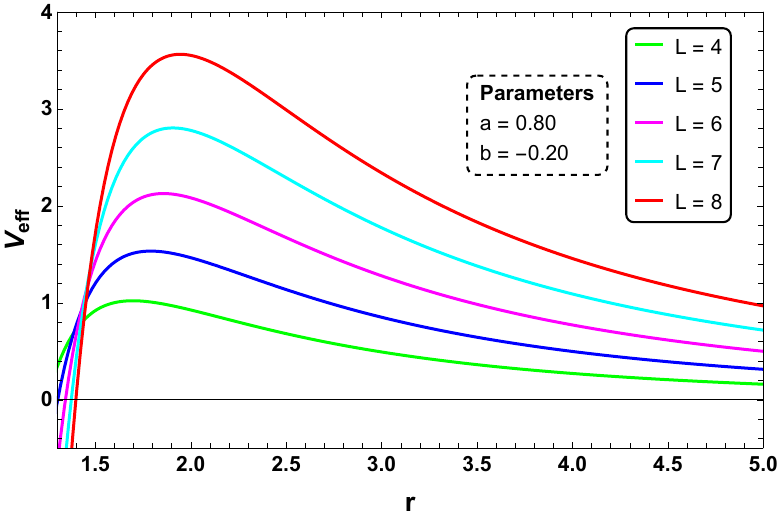}\label{veff_2}}
		\caption{The behaviour of the effective potential $V_{eff}$ versus $r$ at the equatorial plane, showing the influence of the dilaton parameter $b$ and rotation parameter $a$ for different angular momenta $L$. Panel (a) The blue curve corresponds to $L$ within the allowed range; Panel (b) Where $L$ lies outside the allowed range.} \label{eff_emda}
	\end{figure}
	
	\vspace{0.2cm}
	For the time-like particles, $u^{t} \geq 0$ at the equatorial plane, the condition reads as 
	
	\begin{equation}
		E\left(\Xi^{2}-\Delta a^{2}\right) \geq aL\left(\Xi-\Sigma\right),
	\end{equation}
	
	\vspace{0.2cm}
	must be satisfied, as $r$ $\rightarrow$ $r_{H}$, the above condition reduced to
	
	\begin{equation}
		E \geq \frac{a}{\Xi}L = \frac{a}{r_{H}^{2}+2br_{H}+a^{2}}L = \Omega_H L, \label{velo_out}
	\end{equation}
	
	\vspace{0.2cm}
	where $\Omega_H=\frac{a}{\Xi}$ is the angular velocity of the outer event horizon. Moreover, the dependence on the dilaton parameter $b$ is evident from the above equation.
	
	\vspace{0.2cm}
	\noindent
	The behaviour of $\dot{r}$ vs $r$ is illustrated in Fig.~(\ref{rdot_plot}), for various values of the angular momentum $L$, spin parameter $a$, and dilaton parameter $b$. The limiting values of the angular momentum corresponding to freely falling test particles are determined from Eq.~\eqref{l_val} for both the extremal and non-extremal cases of the EMDA black hole, and the results are presented in Tables~(\ref{l_ext}) and~(\ref{l_next}). As shown in the figures, when the angular momentum exceeds the critical value, i.e., $L > L_c$, the geodesic trajectories do not cross the event horizon, and the particles fail to fall into the black hole. Conversely, for $L < L_c$, the geodesics always fall into the black hole, indicating inevitable infall. In the limiting case where $L = L_c\equiv \frac{E}{\Omega_H}$, the particle reaches the event horizon precisely, infall occurs at that boundary. The variation of the effective potential $V_{eff}$ as a function of the radial coordinate $r$ is illustrated in Fig.~(\ref{eff_emda}) for different values of the angular momentum $L$. It is found that when the angular momentum of the particle lies within a specific range, the effective potential becomes negative, indicating that the particle motion is bounded within the gravitational field of the black hole. Conversely, when the angular momentum lies outside this range, the effective potential remains positive for all values of $r$, signifying that the particles are unbounded. 
	
		\vspace{0.3cm}
	\begin{table}[ht!]
		\centering
		\begin{tabular}{|c|c|c|c|c|}
			\hline
			$\bf b$ & $ \bf a_E$ & $\bf r_E$ & $\bf L_2(min)$ & $\bf L_1(max)$ \\
			\hline \hline
			$-0.1$ & $0.9$ & $0.9$ & $-4.82$ & $+3.04$  \\
			$-0.2$ & $0.8$ & $0.8$ & $-4.96$ & $+2.88$  \\
			$-0.3$ & $0.7$ & $0.7$ & $-5.11$ & $+2.71$  \\
			$-0.4$ & $0.6$ & $0.6$ & $-5.25$ & $+2.56$  \\
			$-0.5$ & $0.5$ & $0.5$ & $-5.38$ & $+2.41$  \\
			$-0.6$ & $0.4$ & $0.4$ & $-5.53$ & $+2.27$  \\
			$-0.7$ & $0.3$ & $0.3$ & $-5.67$ & $+2.13$  \\
			$-0.8$ & $0.2$ & $0.2$ & $-5.81$ & $+1.99$  \\
			\hline
		\end{tabular} 
		\caption{The range for the numerical values of max/min angular momentum for
			an extremal rotating EMDA black hole with $M=1$, $M_{0}=1$ and $E=1$.} \label{l_ext}
	\end{table}
	
	\vspace{0.5cm}
	\begin{table}[ht!]
		\centering
		\begin{tabular}{|c|c|c|c|c|c|c|c|}
			\hline
			$\bf b$ & $\bf a$ & $\bf r_-$ & $\bf r_+$ & $\bf L_4(min)$ & $\bf L_3(max)$ \\
			\hline \hline
			$-0.1$ & $0.710000$ & $0.360000$ & $1.440000$ & $-3.876840$ & $2.358947$\\
			$-0.2$ & $0.630000$ & $0.320000$ & $1.280000$ & $-3.059032$ & $1.819677$\\
			$-0.3$ & $0.560000$ & $0.280000$ & $1.120000$ & $-2.219859$ & $1.273286$\\
			$-0.4$ & $0.470000$ & $0.240000$ & $0.960000$ & $-1.329516$ & $0.709839$\\
			\hline
		\end{tabular}
		\caption{The range for the numerical values of max/min angular momentum for
			a non-extremal rotating EMDA black hole with $M=1$, $M_{0}=1$ and $E=1$.}\label{l_next}
	\end{table}
	
	\noindent
	Furthermore, from the Table~(\ref{l_ext}), it is clear that as the dilaton parameter $b$ becomes more negative, both the effective mass and the spin of the EMDA black hole decrease, giving rise to a distinct spin-dilaton coupling. The resultant coupling shifts the event horizon inward and slightly deepens the effective potential well. In the extremal limit, the potential barrier tends to become nearly symmetric around $r = 1 + b$, indicating that capture orbits can occur only for finely tuned values of the angular momentum. Consequently, prograde ($L > 0$) infall becomes increasingly restricted, whereas retrograde ($L < 0$) motion remains comparatively easier.
		
	\paragraph{$\bullet$ Particle collisions in the near-horizon region of an extremal EMDA black hole ---} We aim to investigate the behaviour of $E_{\mathrm{cm}}$ as $r \rightarrow r_{E}$. It is observed that, in this limit, both the numerator and denominator of Eq.~(\ref{Ecm_EMDA}) vanish. Applying l'Hôpital's rule twice, the expression for $E_{\mathrm{cm}}$ as $r \rightarrow r_{E}$ becomes
	
	\begin{align}
		\left. \frac{E_{CM}^2}{2M_0^2} \right|_{r \to r_E}
		=\; C_0 + C_1\big(L_1 + L_2\big) + &C_2\, L_1 L_2 
		+ \frac{G_1^2 (L_2 - L_c)}{8 (L_1 - L_c)^3} 
		+ \frac{G_2^2 (L_1 - L_c)}{8 (L_2 - L_c)^3} \notag \\[10pt]
		&\qquad - \frac{G_1 G_2}{4(L_1 - L_c)(L_2 - L_c)} 
		- \frac{H_1 (L_2 - L_c)}{4 (L_1 - L_c)} 
		- \frac{H_2 (L_1 - L_c)}{4 (L_2 - L_c)}, \label{ecm_ext}
	\end{align}
	
	with coefficient
	
	\[
	C_0 = \frac{26b^2 + 30b + 8}{3b^2 + 4b + 1}, \qquad 
	C_1 = 0, \qquad 
	C_2 = -\frac{1}{3b^2 + 4b + 1},
	\]
	
	\[
	K = 4(1 + b)(1 + 2b), \qquad
	G_i = K\,\bigl(L_c - L_i\bigr), \qquad
	H_i = 40b^2 + 44b + 12 - 2L_i^2, \quad (i = 1, 2).
	\]
	
	\vspace{0.2cm}
	By substituting $b=-0.2$ into the above expression, Eq.~(\ref{ecm_ext}) reads as
	
	\begin{figure}[h!]
		\centering
		\subfigure[]{\includegraphics[width=7.5cm,height=7.8cm]{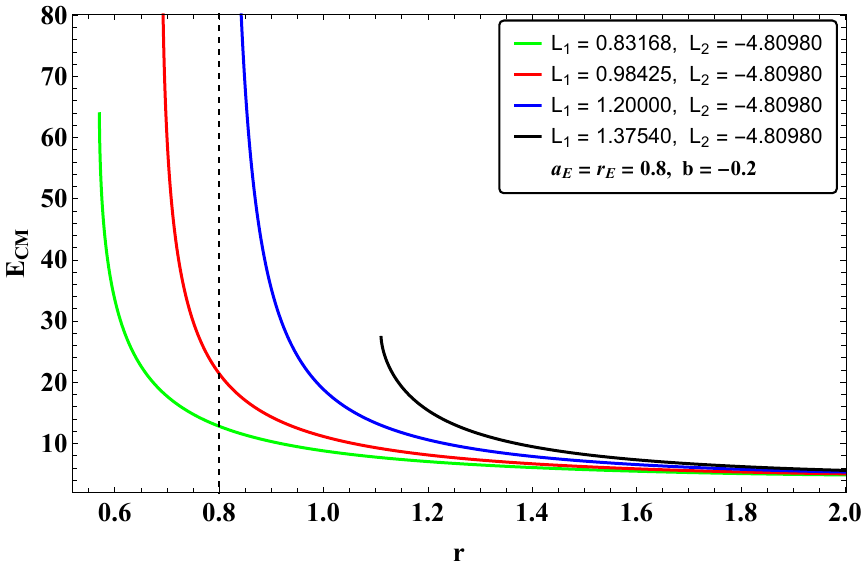}\label{E_cm_r_ext_1}}
		\hspace{0.8cm}
		\subfigure[]{\includegraphics[width=7.5cm,height=7.8cm]{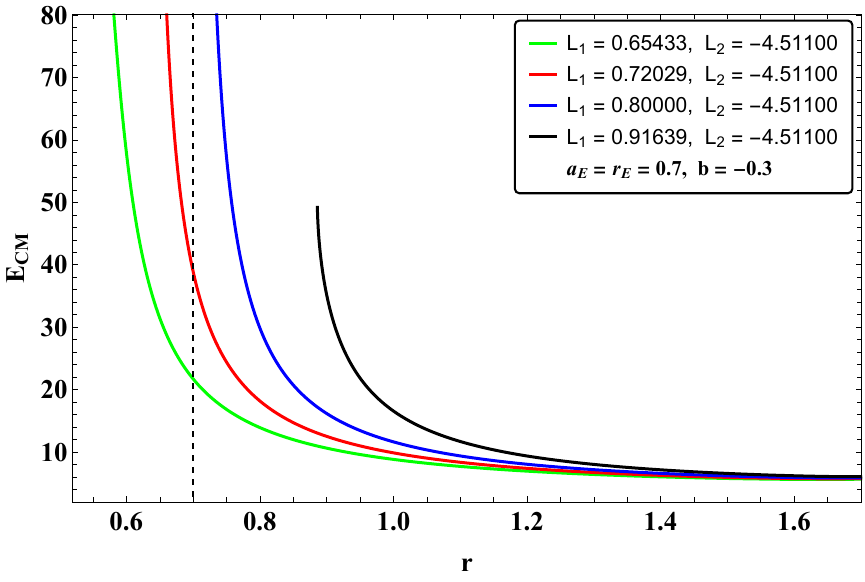}\label{E_cm_r_ext_2}}
		\vspace{0.5em} 
		\subfigure[]{\includegraphics[width=7.6cm,height=7.8cm]{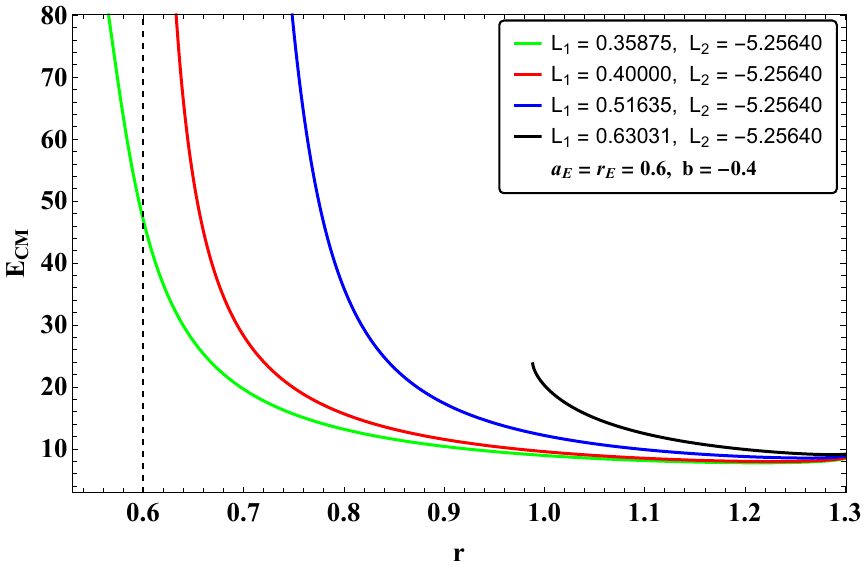}\label{E_cm_r_ext_3}}
		\caption{Variation of the center-of-mass energy $E_{\mathrm{CM}}$ versus $r$ for an extremal EMDA black hole for fixed values of the dilaton parameter $b=-0.2$, $b=-0.3$ and $b=-0.4$, respectively. Each curve corresponds to different values of the angular momentum $L_1$, with $L_2$ held constant. The vertical dashed line in each plot denotes the location of the extremal event horizon $r_E$ of the EMDA black hole.}\label{E_cm_1}
	\end{figure}

	\begin{align}
		\left. \frac{E_{CM}^2}{2M_0^2} \right|_{r \to r_E}
		=\; 9.5 &- 3.123\, L_1 L_2 
		+ \frac{{1.92(L_c-L_1)}^2 (L_2 - L_c)}{8 (L_1 - L_c)^3} 
		+ \frac{{1.92(L_c-L_2)}^2 (L_1 - L_c)}{8 (L_2 - L_c)^3} \notag \\[10pt]
		&- \frac{3.84(L_c-L_1)(L_c-L_2)}{4(L_1 - L_c)(L_2 - L_c)} 
		- \frac{(4.8-2{L_{1}}^2) (L_2 - L_c)}{4 (L_1 - L_c)} 
		- \frac{(4.8-2{L_{2}}^2) (L_1 - L_c)}{4 (L_2 - L_c)}. \label{ecm_ext_1}
	\end{align}
	
	\begin{figure}[h!]
		\centering
		\subfigure[]{\includegraphics[width=7.5cm,height=7.8cm]{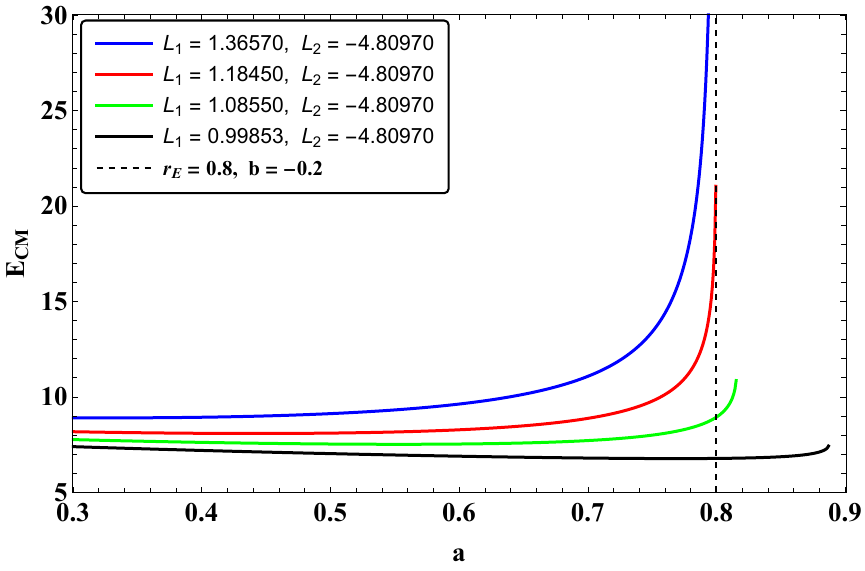}\label{E_cm_r_ext_for_a_fig_1}} \hspace{0.8cm}
		\subfigure[]{\includegraphics[width=7.5cm,height=7.8cm]{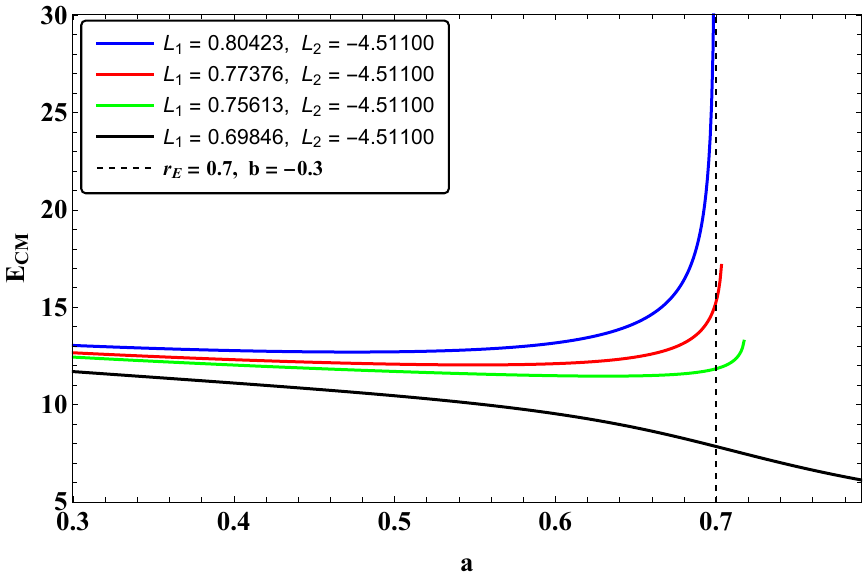}\label{E_cm_r_ext_for_a_fig_2}}
		\caption{Variation of the center-of-mass energy $E_{\mathrm{CM}}$ versus $a$ for an extremal EMDA black hole for fixed values of the dilaton parameter $b=-0.2$ and $b=-0.3$, respectively. Each curve corresponds to different values of the angular momentum $L_1$, with $L_2$ held constant. The vertical dashed line in each plot denotes the location of the extremal event horizon $r_E$ of the EMDA black hole.}\label{E_cm_3}
	\end{figure}
	
	\begin{figure}[h!]
		\begin{centering}
			\subfigure[]{\includegraphics[width=7.5cm,height=7.8cm]{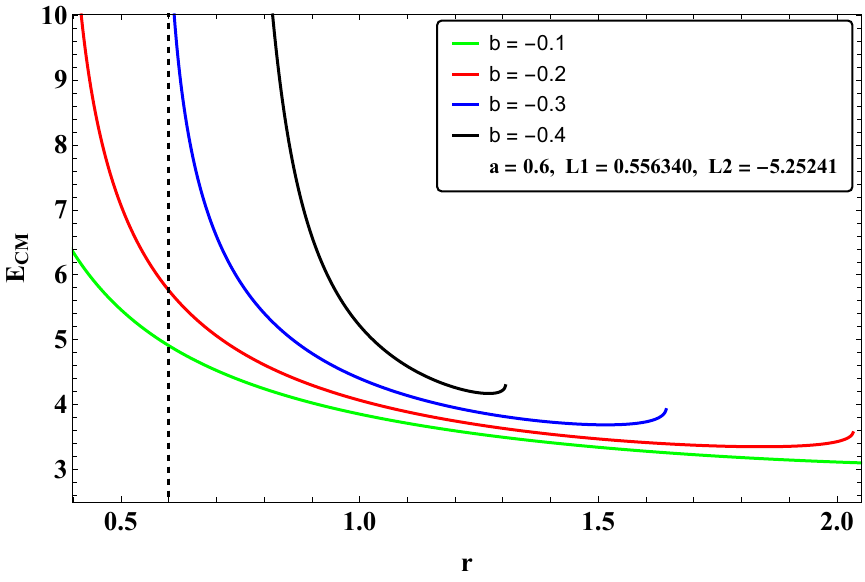}\label{Ecm_ext_a_4}}\hspace{0.8cm}
			\subfigure[]{\includegraphics[width=7.5cm,height=7.8cm]{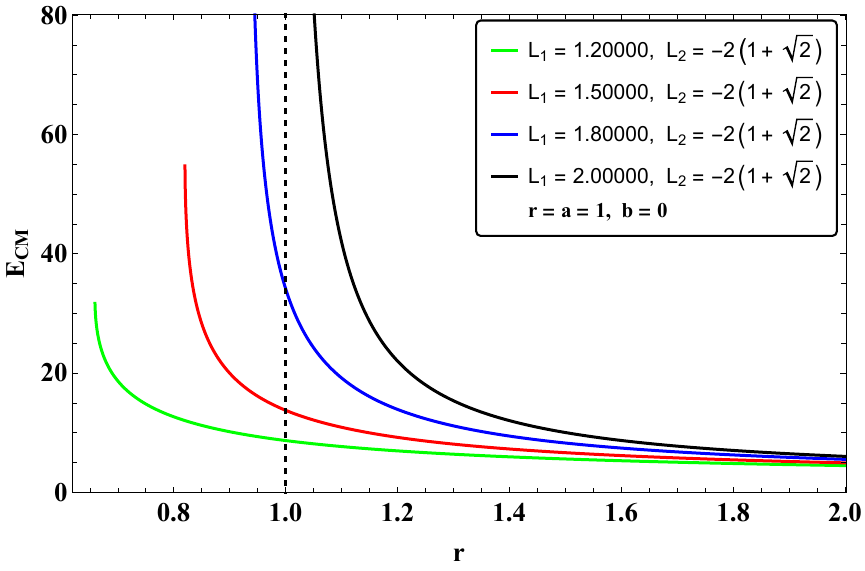}\label{Ecm_ext_a_5}}
			\par\end{centering}
		\caption{Variation of the center-of-mass energy $E_{\mathrm{CM}}$ as a function of the radial coordinate $r$. Panel~(a) shows the extremal EMDA case for different values of the dilaton parameter $b$, while panel~(b) corresponds to the extremal Kerr black hole limit ($b=0$). In both panels, the vertical line indicates the location of the event horizon.}\label{E_cm_2}
	\end{figure}
	
	\begin{figure}[h!]
		\begin{centering}
			\subfigure[]{\includegraphics[width=7.5cm,height=7.8cm]{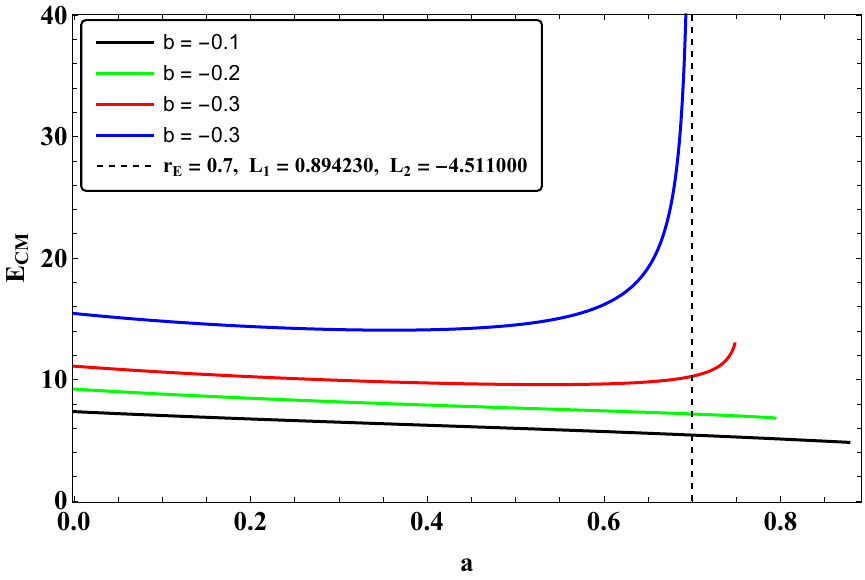}\label{E_cm_next_for_a_fig_3}}\hspace{0.8cm}
			\subfigure[]{\includegraphics[width=7.5cm,height=7.8cm]{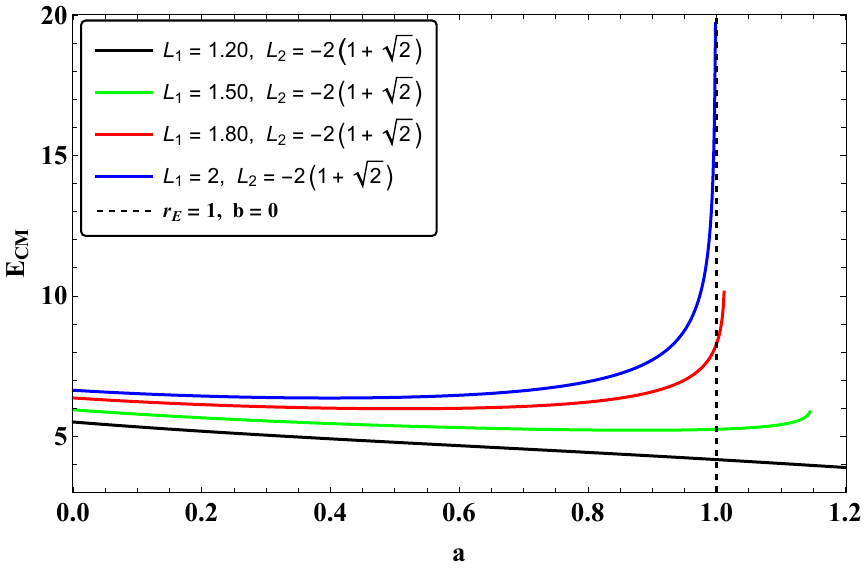}\label{E_cm_next_for_a_fig_4}}
			\par\end{centering}
		\caption{Variation of the center-of-mass energy $E_{\mathrm{CM}}$ versus the spin parameter $a$. Panel~(a) shows the extremal EMDA case for different values of the dilaton parameter $b$, while panel~(b) corresponds to the extremal Kerr black hole limit ($b=0$). In both panels, the vertical line indicates the location of the event horizon.}\label{Ecm_next_plot_a_2}
	\end{figure}
	
	\vspace{0.2cm}
	\noindent
	For $b=-0.2$, one has $a_E=r_E=0.8$ and $L_c=E/\Omega_E=1.2$, as shown in Table~(\ref{l_ext}). Eq.~(\ref{ecm_ext_1}) therefore gives the limiting extremal behaviour of $E_{\rm CM}$ near the horizon when one of the colliding particles approaches the critical angular momentum. The expression shows that the center-of-mass energy $E_{CM}$ diverges (infinity) when either of the angular momenta, $L_{1}$ or $L_{2}$, attains its critical value $L_{c}$. For all other generic values of $L_{1}$ and $L_{2}$, the $E_{CM}$ remains finite. The dilaton parameter $b$ merely shifts the critical parameters $r_E$, $\Omega_E$, $L_c$ and the rate of divergence without destroying the mechanism. This observation implies that an extremal EMDA black hole can, in principle, serve as a natural particle accelerator, capable of accelerating particles to arbitrarily high energies and thereby providing a potential framework to probe Planck-scale physics. However, to achieve an infinite $E_{CM}$, the infalling particles must possess angular momenta within a specific critical range, as shown in Table~(\ref{l_ext}). 
	
	\vspace{0.2cm}
	\noindent
	It is essential at this stage to delineate the precise conceptual and formal relationship between the result derived in the present work and the finite center-of-mass energy reported by Hussain~\citep{2012MPLA...2750068H}. In that study, Hussain examined the radial infall of uncharged particles into an extremal EMDA black hole. Within this restricted framework, azimuthal motion is suppressed, and consequently, no independent conserved angular momentum $L_i$ appears in the final expression for $E_{\rm CM}$. As a result, the BSW critical condition $L_c=E/\Omega_H$ does not emerge in Hussain's formulation. By contrast, the present analysis retains the full equatorial geodesic motion, incorporating nonzero conserved angular momenta $L_1$ and $L_2$ for the two particles. For generic noncritical configurations, where $L_i\neq L_c$, our expression yields a finite horizon value of $E_{\rm CM}$, thereby recovering the qualitative behaviour of Hussain's radial result. The divergence arises exclusively when one of the particles is tuned to the critical angular momentum $L_i=L_c$. Hence, Hussain's finite outcome should be interpreted as a consequence of the restricted radial, noncritical setup, in which the independent angular-momentum degree of freedom is absent. Nevertheless, the present result does not contradict that conclusion; rather, it generalizes the analysis to arbitrary equatorial geodesics and establishes that the BSW divergence is confined to the critical angular-momentum channel $L=L_c$. For further clarity, a detailed near-horizon comparison of the two limiting cases is discussed in Appendix~(\ref{app:append_b}).
	
	\vspace{0.2cm}
	\noindent
	The near-horizon expansion also determines the rate of the center-of-mass energy $E_{CM}$ divergence. Although Eq.~\eqref{Ecm_EMDA} contains an explicit factor $1/\Delta$, the numerator vanishes simultaneously at the extremal horizon. For generic noncritical particles, $L_i\neq L_c$, this cancellation is complete and $E_{CM}$ remains finite. If one particle is tuned to the critical angular momentum $L_i=L_c$, the cancellation is only partial. Since for the	extremal EMDA black hole
		
	\begin{equation}
		a_E=r_E=1+b,\qquad \Delta \sim (r-r_E)^2,
	\end{equation}
	
	\vspace{0.2cm}
	one obtains
	
	\begin{equation}
		\frac{E_{\rm CM}^2}{2M_0^2}
	\sim
	\frac{\mathcal{K}_{\rm EMDA}}{r-r_E}
	=
	\frac{\mathcal{K}_{\rm EMDA}}{\sqrt{\Delta}},
	\qquad
	E_{\rm CM}\sim (r-r_E)^{-1/2}.
	\end{equation}
	
	\vspace{0.2cm}
	\noindent
	Thus the divergence is weaker than a direct $1/\Delta$ pole in $E_{CM}^2$. The parameter $b$ modifies the Kerr result through the shifted extremal horizon $r_E=1+b$, the horizon angular velocity
	
	\begin{equation}
		\Omega_E=\frac{1}{2(1+2b)},
	\end{equation}
	
	\vspace{0.2cm}
	\and therefore the critical angular momentum
	
	\begin{equation}
		L_c=2(1+2b)E_i .
	\end{equation}
	
	\vspace{0.2cm}
	\noindent
	For the case $E_1=E_2=1$, with particle 1 critical ($L_1=L_c$) and particle 2 noncritical ($L_2 \neq L_c$), the coefficient becomes
	
	\begin{equation}
		\mathcal{K}_{\rm EMDA}(b,L_2)
		=
		\frac{(L_c-L_2)
			\left[
			2(1+2b)-\sqrt{2(1+b)(1+2b)}
			\right]}
		{1+3b},
		\qquad
		L_c=2(1+2b).
	\end{equation}

	\vspace{0.2cm}
	\noindent
	In the Kerr limit $b\to0$, this reduces to
	
	\begin{equation}
		\mathcal{K}_{\rm Kerr}(L_2)
		=
		(2-L_2)(2-\sqrt{2}).
	\end{equation}
	
	\vspace{0.2cm}
	Hence the EMDA correction appears through the replacement $L_c=2E_i\rightarrow 2(1+2b)E_i$, together with the additional $b$-dependent geometric prefactor in $\mathcal{K}_{\rm EMDA}$.
	
	\vspace{0.2cm}
	\noindent
	Moreover, Fig.~(\ref{E_cm_1}) illustrates the variation of the $E_{CM}$ as a function of the radial coordinate $r$ for different values of the angular momenta $L_{1}$ and $L_{2}$, keeping the spin parameter $a$ and the dilaton parameter $b$ fixed. It is observed that $E_{CM}$ diverges at the event horizon when either $L_{1}$ or $L_{2}$ attains the critical value $L_{c}$. This indicates that the critical angular momentum $L_{c}$ must lie within the permissible range for which the particle can reach the black hole horizon. In Fig.~(\ref{E_cm_3}), the near-horizon variation of the center-of-mass energy, $E_{\rm CM}$, as a function of the spin parameter $a$ is presented for different choices of $L_{1}$ and $L_{2}$, while the dilaton parameter $b$ is kept fixed. Finally, the dependence of the $E_{CM}$ on the dilaton parameter is illustrated in Fig.~(\ref{Ecm_ext_a_4}), where $E_{\mathrm{CM}}$ is plotted versus the radial coordinate $r$ for different values of $b$, while Fig.~(\ref{E_cm_next_for_a_fig_3}) presents the corresponding variation of $E_{\mathrm{CM}}$ with the spin parameter $a$. For comparison, the extremal Kerr limit ($b=0$) is shown separately in Figs.~(\ref{Ecm_ext_a_5}) and~(\ref{E_cm_next_for_a_fig_4}). Jacobson \citep{PhysRevLett.104.021101} argued that infinite center-of-mass energy can be achieved by colliding particles only after an endless amount of proper time, making such events physically unrealistic. However, in the case of a non-extremal black hole, the proper time required for a particle to reach the event horizon, though significantly large, remains finite. Therefore, it is plausible to study particle collisions near a non-extremal black hole, where such processes could occur within a physically meaningful timescale. In the next paragraph, we will explore the collision of particles near a non-extremal EMDA black hole.
	
	\paragraph{$\bullet$ Particle collisions in the near-horizon region of a non-extremal EMDA black hole ---} Finally, we analyse the behaviour of the center-of-mass energy $E_{CM}$ in the vicinity of the event horizon $r_H^{+}$ for a non-extremal rotating EMDA black hole. As the radial coordinate $r$ approaches $r_H^{+}$, both the numerator and the denominator of Eq.~(\ref{Ecm_EMDA}) simultaneously approach zero. To determine the limiting value $E_{CM}$ near the event horizon under non-extremal conditions, we apply l'Hospital's rule. Finally, the expression for $E_{\mathrm{cm}}$ as $r \rightarrow r_H^{+}$ becomes
	
	\begin{align}
		\frac{E_{CM}^2}{2M_0^2}\Big|_{r\to r_H^+}
		=\frac{1}{K\,(L_3 - L_c')(L_4 - L_c')}
		\Bigg(
		\alpha_0+\alpha_1(L_3+L_4)+\alpha_2(L_3^2+L_4^2)+\alpha_3L_3L_4
		\Bigg), \label{ecm_next}
	\end{align}

	\vspace{0.2cm}
	\text{with coefficients}
	
	\begin{align*}
		K(b,a) &= \frac{a^2}{r_H^{+}(1+2b)}, \\[10pt]
		\alpha_0 &= 8r_H^{+}(1+2b), \qquad
		\alpha_1 = -4a, \\[10pt]
		\alpha_2 &= 1, \qquad
		\alpha_3 = 2\left[
		\frac{a^2}{r_H^{+}(1+2b)}-1
		\right].
	\end{align*}
	
	\vspace{0.2cm}
	\noindent
	By substituting $b=-0.3$, $a=0.56$ and $r_H^+=1.12$ (see Table~\ref{l_next}) into the above expression, Eq.~(\ref{ecm_next}) reads as
	
	\begin{align}
		\frac{E_{CM}^2}{2M_0^2}\Big|_{r\to r_H^+}
		=\frac{1}{0.7\left(L_3-L_c'\right)\left(L_4-L_c'\right)}
		\Bigg(3.584-2.24(L_3+L_4)+L_3^2+L_4^2-0.6L_3L_4\Bigg), \label{ecm_next_2}
	\end{align}
	
	\begin{figure}[h!]
		\begin{centering}
			\subfigure[]{\includegraphics[width=7.5cm,height=7.8cm]{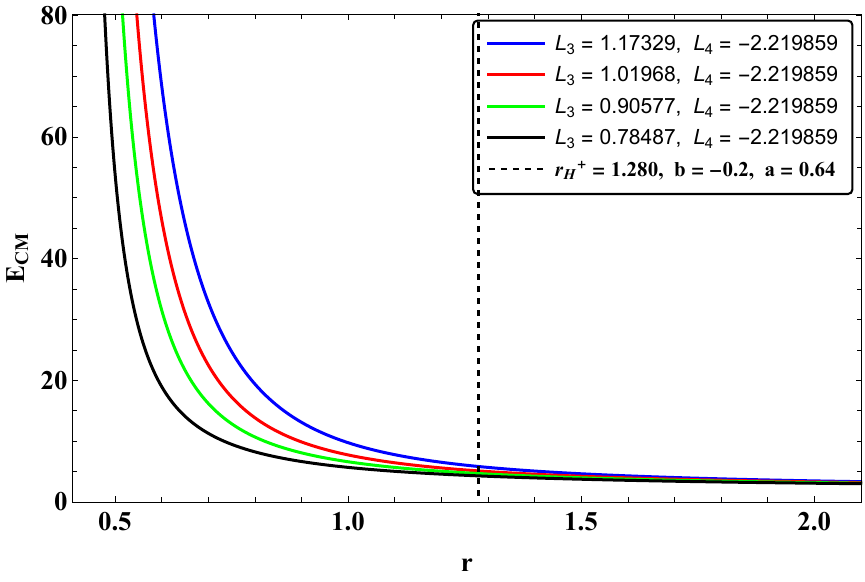}\label{Ecm_next_1}} \hspace{0.8cm}
			\subfigure[]{\includegraphics[width=7.5cm,height=7.8cm]{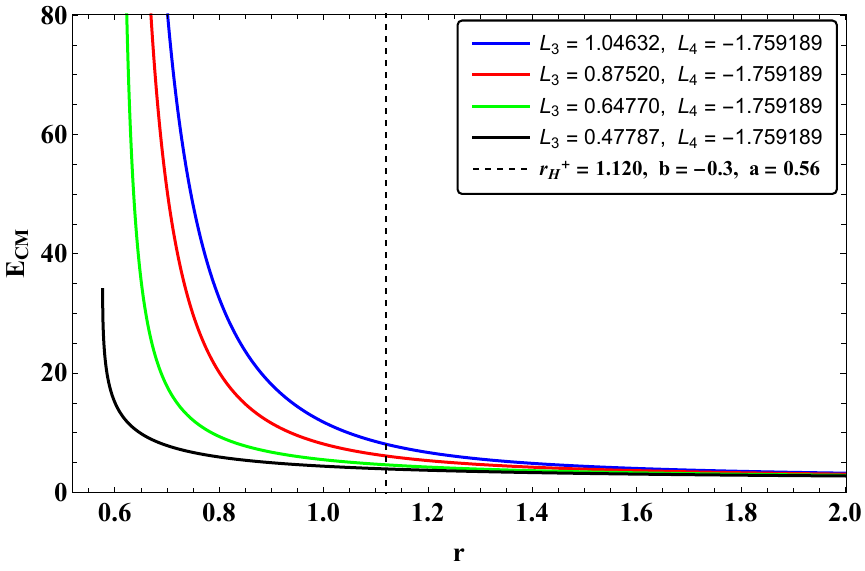}\label{Ecm_next_2}}\vspace{0.8em}
			\subfigure[]{\includegraphics[width=7.5cm,height=7.8cm]{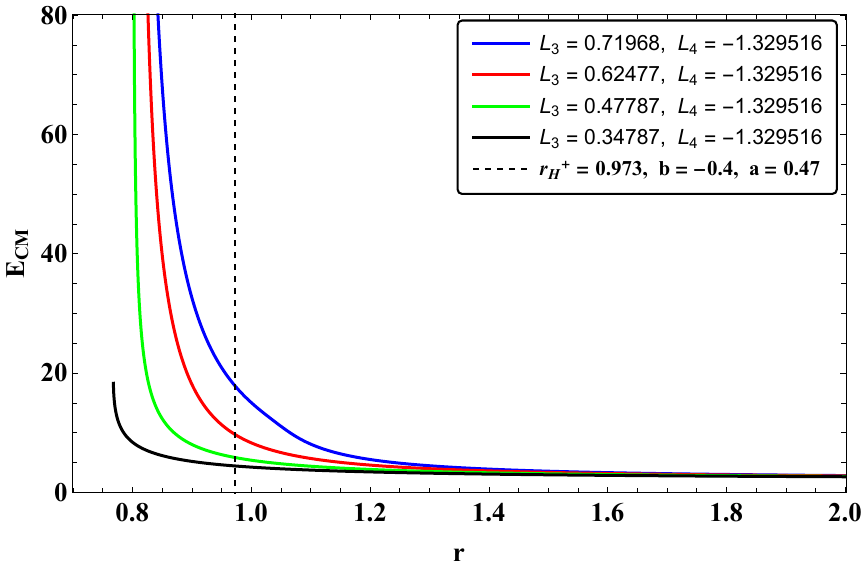}\label{Ecm_next_3}}
			\par\end{centering}
		\caption{Variation of the center-of-mass energy $E_{\mathrm{CM}}$ versus radial coordinate $r$ for a non-extremal EMDA black hole for fixed values of the dilaton parameter $b=-0.2$, $b=-0.3$ and $b=-0.4$, respectively. Each curve corresponds to different values of the angular momentum $L_3$, with $L_4$ held constant. The vertical dashed line in each plot denotes the location of the outer event horizon $r_+$ of the EMDA black hole.} \label{Ecm_next_plot}
	\end{figure}
	
	\vspace{0.2cm}
	\noindent
	where $L_c'=\frac{E}{\Omega_H}=\frac{2r_H^+(1+2b)}{a}=1.60$. The above expression remains invariant under the interchange $L_3 \leftrightarrow L_4$, thereby exhibiting the expected symmetry between the two colliding particles. According to Eq.~(\ref{ecm_next_2}), one might expect that the center-of-mass energy $E_{CM}$ can diverge when either $L_{3}$ or $L_{4}$ equals the critical angular momentum $L'_{c}$. Nevertheless, for such a divergence to occur physically, the particle with angular momentum $L'_{c}$ must be capable of reaching the event horizon. In other words, $L'_{c}$ should lie within the permissible range of angular momenta that allow the particle to fall into the black hole and participate in a collision near the horizon. As shown in Table~(\ref{l_next}), for the parameter set $b=-0.3$ and $a=0.56$, the allowed range of angular momentum is $L_4=L_{\min}<L<L_{\max}=L_3$. Hence	the critical angular momentum $L_c'$ does not lie within this interval; rather, it exceeds the upper bound, satisfying $L_c'>L_3=L_{\max}$. This indicates that, in the case of a non-extremal EMDA black hole, a particle with angular momentum $L = L'_{c}$ is unable to reach the event horizon or fall into the black hole. Consequently, $E_{CM}$ for the non-extremal configuration possesses a finite maximum value. Fig.~(\ref{Ecm_next_plot}) illustrates the near-horizon variation of the center-of-mass energy, $E_{\rm CM}$, as a function of the radial coordinate $r$ for different choices of $L_{3}$ and $L_{4}$, with the parameters $a$ and $b$ fixed at selected combinations, for a non-extremal EMDA black hole. 
	
	\begin{figure}[h!]
		\begin{centering}
			\subfigure[]{\includegraphics[width=7.5cm,height=7.8cm]{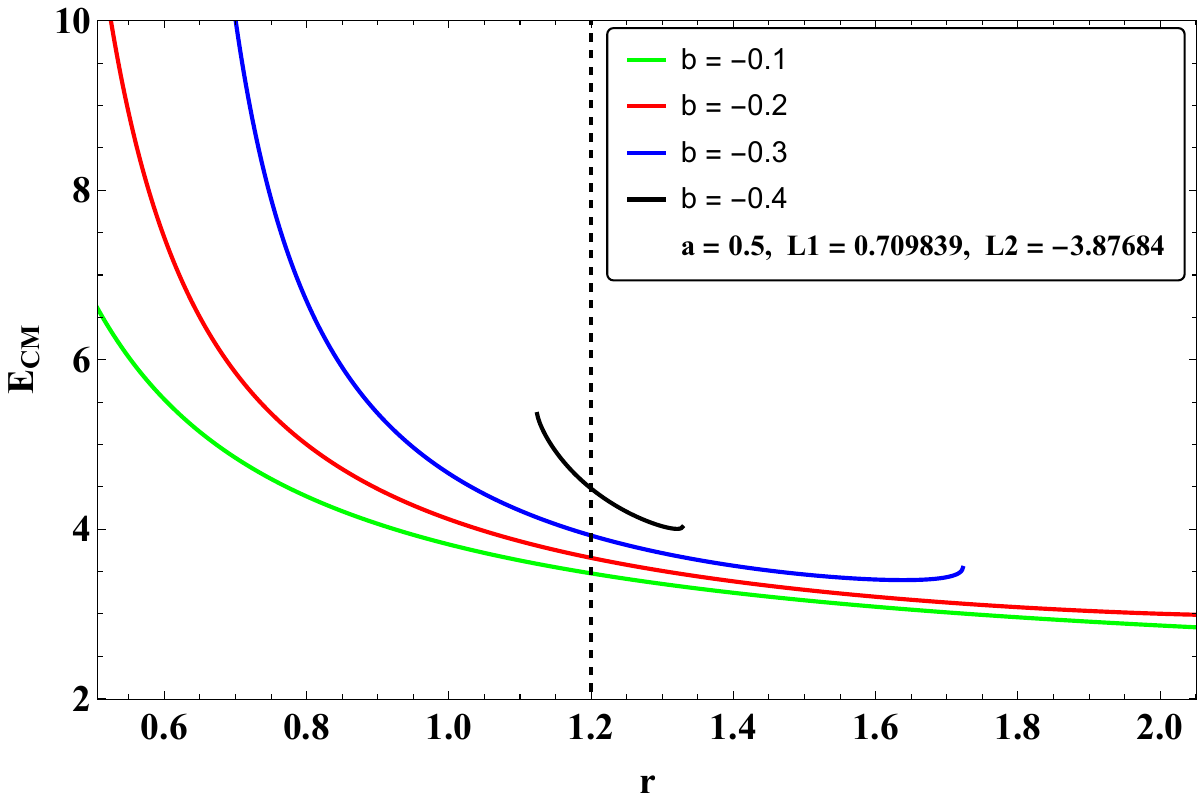}\label{Ecm_next_4}}\hspace{0.8cm}
			\subfigure[]{\includegraphics[width=7.5cm,height=7.8cm]{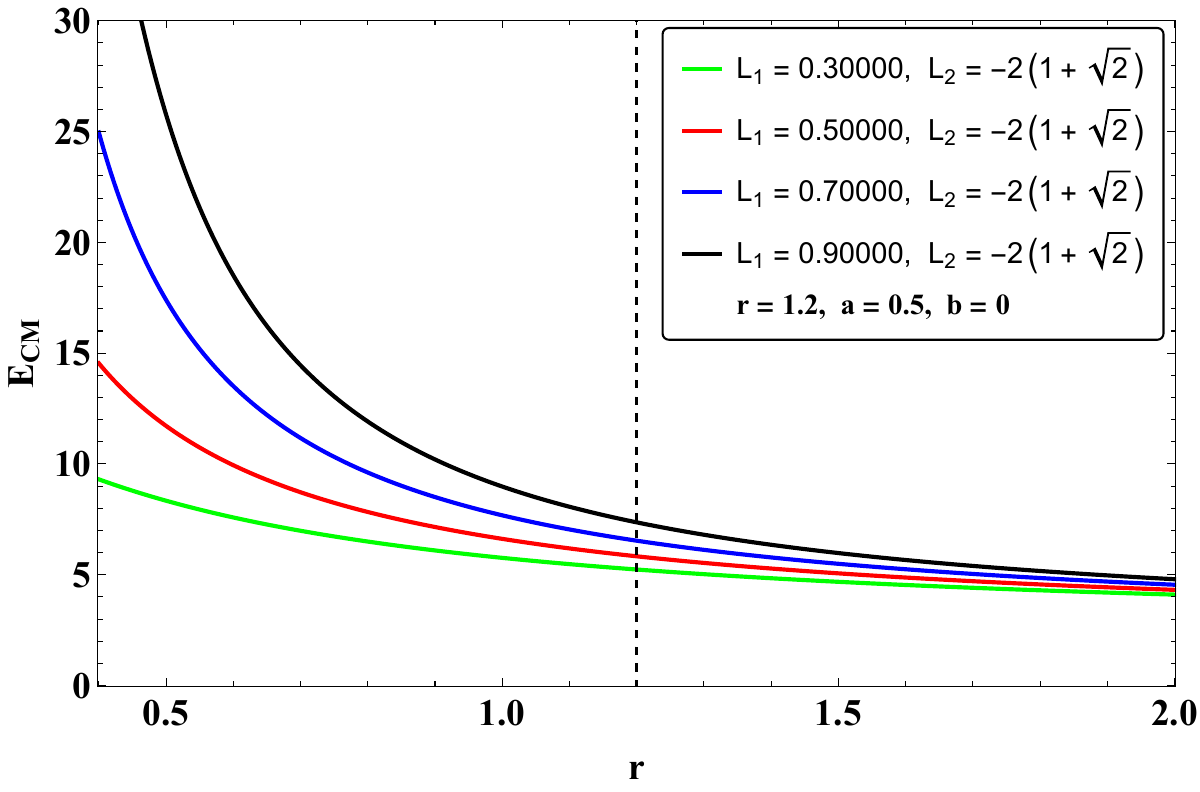}\label{Ecm_next_5}}
			\par\end{centering}
		\caption{Variation of the center-of-mass energy $E_{\mathrm{CM}}$ as a function of the radial coordinate $r$. Panel~(a) shows the non-extremal EMDA case for different values of the dilaton parameter $b$, while panel~(b) corresponds to the non-extremal Kerr black hole limit ($b=0$). In both panels, the vertical line indicates the location of the event horizon.}\label{E_cm_5}
	\end{figure}
	
	\begin{figure}[h!]
		\begin{centering}
			\subfigure[]{\includegraphics[width=7.5cm,height=7.8cm]{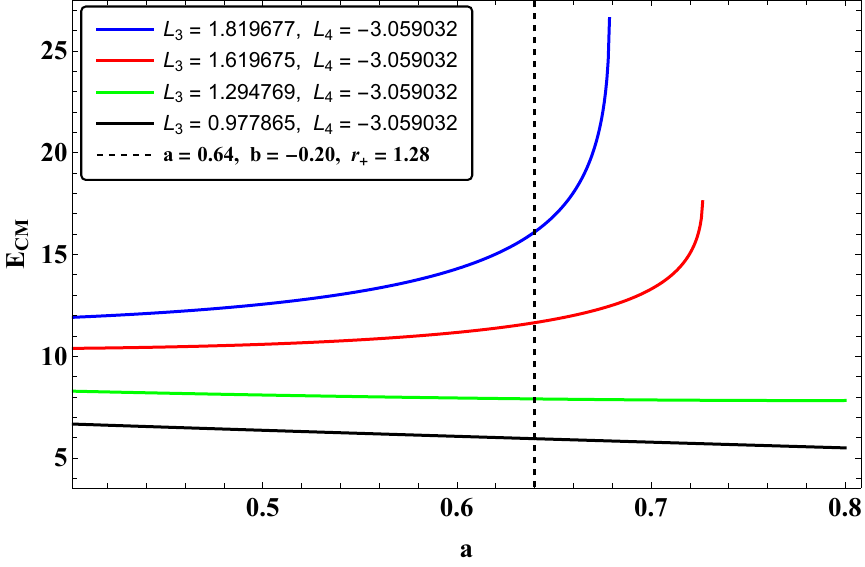}\label{E_cm_next_for_a_fig_1}} \hspace{0.8cm}
			\subfigure[]{\includegraphics[width=7.5cm,height=7.8cm]{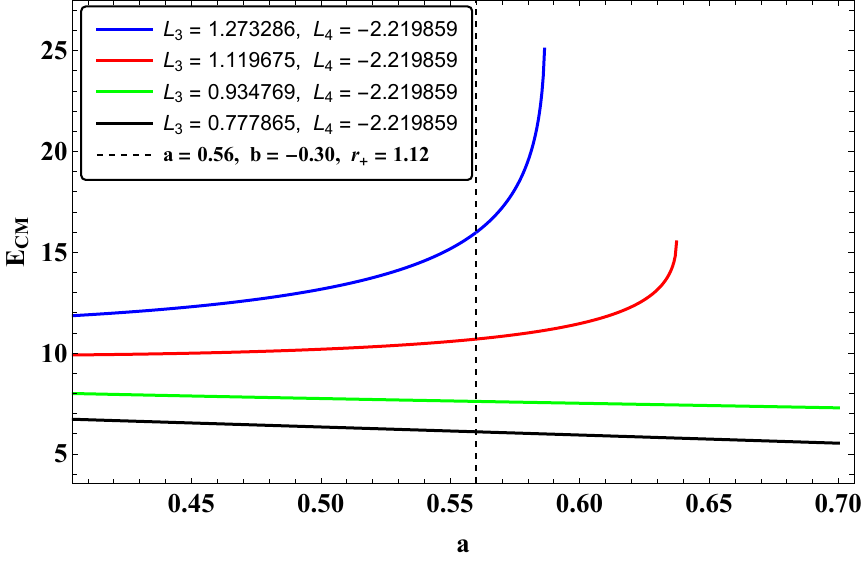}\label{E_cm_next_for_a_fig_2}}
			\par\end{centering}
		\caption{Variation of the center-of-mass energy $E_{\mathrm{CM}}$ versus the spin parameter $a$ for a non-extremal EMDA black hole for fixed values of the dilaton parameter $b=-0.2$ and $b=-0.3$, respectively. Each curve corresponds to different values of the angular momentum $L_3$, with $L_4$ held constant. The vertical dashed line in each plot denotes the location of the outer event horizon $r_+$ of the EMDA black hole.} \label{Ecm_next_plot_a_1}
	\end{figure}
	
	\vspace{0.2cm}
	\noindent
	The finite character of the $E_{CM}$ in the near-horizon region of non-extremal EMDA black hole can be compared directly with the Kerr case by taking $b=0$. In this limit,
	
	\begin{equation}
		r_H^+ \rightarrow r_+^{\rm Kerr}=1+\sqrt{1-a^2},
		\qquad
		\Omega_H \rightarrow \Omega_H^{\rm Kerr}
		=
		\frac{a}{(r_+^{\rm Kerr})^2+a^2},
	\end{equation}
	
	\vspace{0.2cm}
	and hence
		
	\begin{equation}
		L_c' \rightarrow L_{c,{\rm Kerr}}'
		=
		\frac{1}{\Omega_H^{\rm Kerr}} .
	\end{equation}

	\vspace{0.2cm}
	\noindent
	Therefore, the EMDA expression reduces smoothly to the standard non-extremal Kerr form. For the representative EMDA configuration $b=-0.3$, $a=0.56$, and
	$r_H^+=1.12$, one obtains
	
	\begin{equation}
		\Omega_H^{\rm EMDA}=0.625,
		\qquad
		L_c'=1.60.
	\end{equation}
	
	\vspace{0.2cm}
	\noindent
	However, the allowed angular-momentum range for particles reaching the horizon
	is
	
	\begin{equation}
			L_{\min}<L<L_{\max},
		\qquad
		L_{\min}=-1.759189,\quad L_{\max}=1.04632 .
	\end{equation}
	
	\vspace{0.2cm}
	Therefore,
	
	\begin{equation}
		L_c'>L_{\max},
	\end{equation}

	\vspace{0.2cm}
	\noindent
	so the critical particle cannot reach the horizon and the apparent pole in the
	non-extremal expression is not physically accessible. For comparison, a non-extremal Kerr black hole with the same spin $a=0.56$ has
	
	\begin{equation}
			r_+^{\rm Kerr}=1+\sqrt{1-a^2}=1.82849,
		\qquad
		\Omega_H^{\rm Kerr}=0.15311,
		\qquad
		L_{c,{\rm Kerr}}'=6.531.
	\end{equation}
	
	\vspace{0.2cm}
	\noindent
	This critical value also lies outside the allowed infall range for particles
	released from rest at infinity. Hence both non-extremal Kerr and non-extremal
	EMDA black holes yield finite $E_{\rm CM}$ at the outer horizon. The role of
	the dilaton parameter $b$ is to shift the horizon radius, increase
	$\Omega_H$, and lower the formal critical angular momentum, but not to make
	the non-extremal BSW pole physically accessible. Fig.~(\ref{E_cm_5}) further illustrates the influence of the dilaton parameter on the center-of-mass energy in the non-extremal EMDA geometry, with the non-extremal Kerr spacetime included as the limiting case for direct comparison. Finally, Fig.~(\ref{Ecm_next_plot_a_1}) presents the corresponding dependence of $E_{\rm CM}$ on the spin parameter $a$ for different values of $L_{3}$ and $L_{4}$, with the dilaton parameter $b$ kept fixed.
	
	\section{Discussion \& Summary}\label{sec_7}
	Classical general relativity predicts the formation of spacetime singularities
	through the singularity theorems. Such singularities are generally interpreted
	as signals of the breakdown of the classical theory rather than as physically
	complete descriptions of nature. In the absence of a definitive theory of
	quantum gravity, phenomenological extensions of general relativity and
	modified black hole models provide useful frameworks for probing the
	singularity problem and possible deviations from the Kerr paradigm. Although
	the no-hair theorem suggests that astrophysical black holes in general
	relativity should be described by the Kerr solution, whether this idealised
	description fully captures real compact objects remains an open question.
	Accordingly, black holes beyond the Kerr family are often studied as
	theoretical laboratories for testing strong-field gravity.
	
	\vspace{0.2cm}
	\noindent
	In the present work, we investigate the rotating EMDA black hole, focusing in
	particular on the role of the dilaton parameter $b$. This parameter modifies
	the horizon structure, ergoregion, near-horizon geometry, and particle
	dynamics, thereby providing a clear route for distinguishing EMDA black holes
	from their Kerr counterparts. We analyse the horizon properties and examine
	energy extraction through the Penrose process, showing how $b$ affects the
	maximum extraction efficiency. To quantify the rotational energy available
	from the black hole itself, we further compute the irreducible mass and the
	corresponding extractable rotational energy. The kinematic viability of the
	Penrose process is then assessed using the Wald and	BPT bounds, which determine the minimum local velocities required for fragment production. Finally, we study scalar-field superradiance by deriving the horizon flux and the superradiant condition, thereby clarifying how the dilaton parameter influences the angular velocity of the horizon and the negative horizon-flux regime. The results obtained in the present work should be interpreted within the idealised framework of the test-particle and test-field approximations. In particular, the largest Penrose process efficiencies and the BSW-type enhancement of the center-of-mass energy occur in the extremal or near-extremal regime. Exact extremality is not expected to be realised in realistic astrophysical black holes, and any deviation from extremality regularises the BSW divergence, so that $E_{\rm CM}$ remains large but finite. Hence the formally divergent collision energy should be viewed as a probe of	the near-horizon kinematic structure of the EMDA geometry, rather than as a direct prediction of infinite-energy particle collisions. Moreover, backreaction, self-force effects, radiation reaction, finite-size corrections, and possible quantum-gravitational effects have been neglected in the present analysis. These effects are expected to lower the attainable extraction efficiencies and collision energies. The calculations have also been performed in a vacuum EMDA background. In realistic environments, accretion plasma, magnetic fields, radiative losses, turbulence, and matter interactions may alter geodesic motion, the formation of negative-energy states, the superradiant boundary conditions, and the escape of energetic fragments or collision products to infinity. Therefore, the efficiencies, superradiant windows, and ultra-high-energy collision scales reported here should be regarded
	as theoretical upper bounds and geometric benchmarks. They quantify how the
	dilaton parameter $b$ modifies the ideal Kerr-like limits of energy extraction
	and particle acceleration, rather than providing direct astrophysical
	predictions.
	
	\vspace{0.2cm}
	\noindent
	The essential conclusions emerging from our investigations can be outlined as follows.
	
	\begin{itemize}
		
		\item [(i)] The structure of the horizon for this nontrivial spacetime has been analysed to determine the critical (or extremal) values of the parameters $a = a_E$ and $b$, which characterise the condition for the existence of an extremal black hole with a degenerate horizon. These extremal values define a boundary that clearly separates the black hole regime from that of a naked singularity in the parameter space. In addition, we have also examined the allowed range of the parameter $b$ corresponding to physically meaningful (positive) values of the radial coordinate $r$. Finally, we extended our analysis to illustrate the boundary that distinguishes a black hole from a naked singularity using the $b$ - $a$ parameter space diagram.
		
		\item [(ii)] The dilaton parameter $b$ has no direct observational measurement at present in the EMDA framework. Current black hole observations are usually reported as constraints on deviations from the Kerr spacetime rather than as direct bounds on a specific EMDA dilaton charge. Thus, the values of $b$ used here should be viewed as a controlled theoretical upper-bound of string-inspired departures from Kerr. Observationally viable values would have to be inferred
		from a dedicated comparison with black hole shadow, accretion, and
		gravitational-wave data, including degeneracies with spin, inclination,
		plasma physics, and environmental effects.
		
		\item [(iii)] The dilaton parameter $b$ is constrained by the requirement that the rotating EMDA spacetime represents a regular black hole exterior. In the present normalisation $M=1$, real electric charge requires $b\leq0$, while the area-positive black hole domain requires $b>-1/2$. In the extremal branch, however, a stronger condition follows from demanding that the positive equatorial curvature singularity remains hidden inside the event horizon. This gives the regular exterior bound $-\frac{1}{3}<b\leq0$. Therefore, the results obtained for negative $b$, including the enhanced Penrose efficiency, enlarged superradiant window, and BSW behaviour, should be understood only within this admissible range and should not be extrapolated beyond the limiting value $b=-1/3$.
		
		\item [(iv)] The equations of motion governing the energy extraction processes have been derived and analysed accordingly. The most striking results emerge when the parameter $b$ takes on increasingly negative values in the case of an extremal black hole. Under such conditions, the Penrose process is found to be significantly more efficient, reaching up to approximately $91\%$ for the illustrative extremal EMDA configuration with $b=-0.3$, compared with the standard extremal Kerr benchmark of approximately $20.7\%$. This enhanced value should be understood as a model-dependent theoretical upper bound within the test-particle approximation rather than as a universal maximum for rotating black holes. Furthermore, it is shown that through the Penrose process, it is possible to extract nearly $62.6\%$ of the initial mass of an extremal black hole in this scenario. These figures, however, represent idealised reversible upper bounds within the test-particle approximation and should not be viewed as directly realisable astrophysical predictions.
		
		\item [(v)] Next, we examined the Wald inequality for this spacetime. Unlike the Kerr black hole, the EMDA extremal EMDA black hole satisfies $a_E=r_E=1+b$, so both the extremal spin and the horizon radius decrease linearly with negative $b$. Kinematically, the Wald bound on the local fragment speed at the horizon is relaxed in the case of EMDA spacetime, i.e., $\lvert v\rvert_{\min}$ decreases monotonically as $b$ becomes more negative—by $\sim\!6.5\%$ at $b=-0.1$, $\sim\!18.4\%$ at $b=-0.2$, and 50\% at $b=-0.3$ relative to Kerr case. Consequently, although EMDA spacetime tightens the extremal spin ceiling ($a_E=1+b$), it simultaneously lowers the threshold for horizon crossing negative-energy influx set by Wald, rendering energy extraction processes more permissive for the same mass scale.
		
		\item[(vi)] Furthermore, with the Bardeen-Press-Teukolsky inequality, we find that a negative dilaton parameter $b$ linearly reduces the extremal spin and horizon radius ($a_E=r_E=1+b$), and concomitantly lowers the near-horizon kinematic thresholds relative to Kerr, i.e., $\lvert v\rvert_{\min}$ decreases by $\sim$8.8\% ($b=-0.1$), 23.3\% ($b=-0.2$), and 55.9\% ($b=-0.3$), while $\lvert v\rvert_{\text{rel}}$ drops by $\sim$6.5\%, 18.4\%, and 50.0\%, respectively. Finally, EMDA black hole enforces a tighter global spin bound (smaller $a_{\max}$) yet a more permissive local Wald threshold, thereby facilitating horizon-crossing energy extraction despite a reduced extremal parameter space.
		
		\item [(vii)] The superradiant scattering phenomenon in the EMDA spacetime is significantly stronger than in the Kerr geometry due to the influence of the dilaton charge $b$. Negative values of $b$ reduce the horizon radius and enhance $\Omega_H$, thereby widening the superradiant frequency window $0 < \beta < k\Omega_H$ and deepening the energy-flux minimum. This enhancement is primarily driven by the increase in $\Omega_H$ resulting from the reduction in $\Xi_H$, rather than by modifications to the effective potential. We emphasise that the present analysis does not compute the full greybody factor; superradiance is inferred from the sign of the horizon flux. Our conclusions therefore refer to the widened superradiant window and negative flux condition, not to a quantitative amplification factor.
		
		\item [(viii)] We compute the effective potential and, by imposing the circular-geodesic conditions $\dot r=0$ and $\partial_r V_{\rm eff}=0$ (with stability assessed via $\partial_r^{2} V_{\rm eff}$), determine the corresponding range of angular momentum. Further, from the behaviour of $\dot r$ as a function of $r$, we identify the critical angular momentum $L_c$ for which the radial motion remains allowed all the way down to the event horizon, ensuring that the colliding particles can indeed reach the black hole horizon. Lastly, using these conserved quantities, we compute the center-of-mass energy $E_{\rm CM}$ of the collision.
		
		\item [(ix)] We have analysed the center-of-mass energy $E_{CM}$ of particle collisions near rotating EMDA black holes, explicitly demonstrating how the dilaton parameter $b$ governs the collision energetics. It is important to clarify that Hussain's finite result for the extremal EMDA black hole is recovered as a special case of our general equatorial expression when the angular momenta of the colliding particles are set to zero, $L_1 = L_2 = 0$, yielding $E_{\rm CM} = 2M_0$. In the more general setting with nonzero angular momenta, the BSW mechanism is found to depend simultaneously on the rotation parameter $a$ and the dilaton parameter $b$. While the formalism smoothly reduces to the Kerr black hole case in the limit $b\to 0$, negative values of $b$ modify the horizon structure, angular velocity, and critical angular momentum without destroying the BSW mechanism. For extremal EMDA black holes, $E_{CM}$ diverges when one of the colliding particles attains the critical angular momentum $L_i = L_c = E_i/\Omega_H$, indicating that such spacetimes can, in principle, act as natural particle accelerators. Thus, Hussain's result and the BSW divergence correspond to distinct noncritical and critical angular-momentum sectors of the same geodesic family. In contrast, for non-extremal EMDA black holes, the critical angular momentum lies outside the physically admissible range, ensuring that the center-of-mass energy remains finite. These results establish the dilaton field as a key regulator of high-energy collision processes in EMDA spacetimes, clarifying the conditions under which ultra-high-energy particle interactions may arise in strong-gravity regimes.
		
		\item [(x)] We finally emphasize that the efficiencies and center-of-mass energies obtained in this present work should be interpreted as idealized geometric upper bounds. The analysis is based on neutral test-particle motion in a stationary and vacuum EMDA geometry, with no backreaction, radiation reaction, plasma effects, magnetic fields, or quantum corrections included. Exact extremality is also a limiting mathematical configuration, and realistic accreting black holes are expected to remain below this limit. Hence, the large value of $\eta_{\max}$ and the formal BSW divergence should not be viewed as direct astrophysical predictions. Instead, they quantify how the dilaton parameter $b$ reshapes the near-horizon kinematics relative to Kerr and other Kerr family black holes. In realistic environments, accretion plasma, magnetic fields, finite-size effects, self-force corrections, and radiative losses are expected to reduce the net efficiency and regularize the formally divergent collision energy.
		
	\end{itemize}

\section*{Acknowledgments}
	
AKC is grateful to the anonymous referees for the constructive comments and suggestions, which significantly improved the clarity and presentation of the present work. AKC also thanks Parthapratim Pradhan for valuable discussions on energy extraction and particle acceleration in general relativity, particularly concerning the geometrical formulation of the Penrose process, and for his comments on earlier versions of the manuscript. The author further acknowledges Hemwati Nandan and Pankaj Sheoran for their insightful suggestions, as well as the facilities provided by ICARD, Gurukula Kangri (Deemed to be University), Haridwar, India.

\appendix

\appendix

\section*{Appendix}

\refstepcounter{section}
\subsection*{Appendix \thesection: Derivation of the Algebraic Identity for Negative-Energy States}
\label{app:append_a}
\addcontentsline{toc}{section}{Appendix \thesection: Derivation of the Algebraic Identity for Negative-Energy States}

In this Appendix, we explicitly derive the algebraic identity associated with the negative-energy states in the EMDA spacetime. We emphasise that Eq.~\eqref{inden} is not obtained by assuming a Kerr-like form or through a term-by-term comparison with the Kerr spacetime. Rather, the identity emerges naturally in the course of simplifying the condition required for the existence of negative-energy states. In particular, starting from the negative-energy condition obtained from the positive branch of Eq.~\eqref{sup13}, we have

\begin{equation}
	(\Xi-\Delta)^2a^2L^2
	>
	\Delta\left[
	(a^2-\Xi)^2L^2
	+
	(\Xi^2-a^2\Delta)(Q+m_0^2(r^2+2br))
	\right],
\end{equation}

\vspace{0.2cm}
\noindent
Collecting the terms proportional to $L^2$, we get

\begin{equation}
	\left[
	a^2(\Xi-\Delta)^2
	-\Delta(a^2-\Xi)^2
	\right]L^2
	>
	\Delta(\Xi^2-a^2\Delta)(Q+m_0^2(r^2+2br)).
\end{equation}

\vspace{0.2cm}
\noindent
The algebraic combination appearing on the left-hand side can be factorised in the following form:

	\begin{align}
		a^2(\Xi-\Delta)^2
		-\Delta(a^2-\Xi)^2
		&=
		a^2(\Xi^2-2\Xi\Delta+\Delta^2)
		-\Delta(a^4-2a^2\Xi+\Xi^2)
		\nonumber\\[12pt]
		&=
		a^2\Xi^2+a^2\Delta^2-a^4\Delta-\Delta\Xi^2
		\nonumber\\[12pt]
		&=-
		(\Xi^2-a^2\Delta)(\Delta-a^2).
	\end{align}

\vspace{0.2cm}
\noindent
Thus, the combination $(\Xi^2-a^2\Delta)(\Delta-a^2)$ arises naturally, with an overall minus sign, in the course of simplifying the negative-energy condition. Upon substituting the explicit forms of the metric functions, we obtain

\begin{equation}
	\Xi=r^2+2br+a^2,
	\qquad
	\Delta=r^2-2(M+b)r+a^2,
\end{equation}

\vspace{0.2cm}
\noindent
Hence we get

\begin{equation}
	\Delta-a^2 = r(r-2M-2b),
\end{equation}

and

\begin{align}
	\Xi^2-a^2\Delta
	&=
	(r^2+2br+a^2)^2
	-a^2\left[r^2-2(M+b)r+a^2\right]
	\nonumber\\[12pt]
	&=
	r^4+4br^3+(a^2+4b^2)r^2
	+2a^2(M+3b)r
	\nonumber\\[12pt]
	&=
	r\left[
	r(r+2b)^2+a^2(r+2M+6b)
	\right].
\end{align}

\vspace{0.2cm}
\noindent
Therefore,

\begin{equation}
	(\Xi^2-a^2\Delta)(\Delta-a^2) =	r^2(r-2M-2b)\left[r(r+2b)^2+a^2(r+2M+6b)\right],
\end{equation}

\vspace{0.2cm}
\noindent
which corresponds to Eq.~\eqref{inden} presented in Section~\ref{sec_4}.

\refstepcounter{section}
\subsection*{Appendix \thesection: Relation with Hussain's radial EMDA result}
\label{app:append_b}
\addcontentsline{toc}{section}{Appendix \thesection: Relation with Hussain's radial EMDA result}

In this appendix we will explicitly demonstrate the relation between the finite center-of-mass energy reported by Hussain~\citep{2012MPLA...2750068H} and the BSW-type divergence obtained in the present work. Although the structural notation used by Hussain differs from that of the present work, the central feature of his result is unambiguous: in the strictly radial sector considered there, the center-of-mass energy of the colliding particles remains finite at the extremal EMDA horizon. The center-of-mass energy of two colliding particle is expressed as

\begin{equation}
	E_{\rm CM}^2
	=
	-g_{\mu\nu}
	\left(P_1^\mu+P_2^\mu\right)
	\left(P_1^\nu+P_2^\nu\right),
	\qquad
	\frac{E_{\rm CM}^2}{2M_0^2}
	=
	1-g_{\mu\nu}u_{(1)}^\mu u_{(2)}^\nu .
\end{equation}

\vspace{0.2cm}
\noindent
Hussain considered two uncharged particles falling radially in the extremal
EMDA geometry. In that restricted sector one imposes

\begin{equation}
	\dot{\theta}=0,\qquad \dot{\phi}=0,\qquad q_1=q_2=0 ,
\end{equation}

\vspace{0.2cm}
\noindent
so that no independent conserved angular momentum $L_i$ appears in the final expression. It should be noted that the original BSW analysis also considered particles falling radially towards the horizon, but with nonzero conserved angular momenta $L_i$, $i=1,2$. Thus, radial infall in the BSW sense does not imply the strictly zero-angular-momentum sector $L_i=0$. Using Hussain's radial four-velocities, the center-of-mass energy takes the form

\begin{equation}
	E_{\rm CM}
	=
	\sqrt{2}\,m
	\left[
	1+
	\frac{\Delta E_1E_2}{\Sigma-a^2}
	-
	\frac{
		\sqrt{
			\left(\Delta E_1^2-2(\Sigma-a^2)\right)
			\left(\Delta E_2^2-2(\Sigma-a^2)\right)}
	}{\Sigma-a^2}
	\right]^{1/2}.
\end{equation}

\vspace{0.2cm}
\noindent
In the extremal limit $r_+=a$, this gives Hussain's finite horizon result,

\begin{equation}
	E_{\rm CM}
	=
	\sqrt{2}\,m
	\left[
	1+
	\frac{E_1E_2(a-2D)}{a-2M}
	+
	\frac{
		\sqrt{
			\left(E_1^2(a-2D)-(a-2M)\right)
			\left(E_2^2(a-2D)-(a-2M)\right)}
	}{a-2M}
	\right]^{1/2}.
	\label{app_hussain_ext}
\end{equation}

\vspace{0.2cm}
\noindent
Hence Hussain's result corresponds to a finite radial, noncritical sector. We now recover the same noncritical limit from our general equatorial expression obtained in Section~(\ref{sec_6}). For $E_1=E_2=1$, our result can be written as

\begin{equation}
	\frac{E_{\rm CM}^2}{2M_0^2}
	=
	\frac{\mathcal{N}(r,a,b,L_1,L_2)}
	{r(r+2b)\left[a^2+r\{r-2(1+b)\}\right]} .
\end{equation}

\vspace{0.3cm}
\noindent
The radial/noncritical limit corresponding to Hussain's sector is obtained by setting

\begin{equation}
	L_1=L_2=0 .
\end{equation}

\vspace{0.2cm}
In this limit,

\begin{equation}
	\mathcal{N}\big|_{L_1=L_2=0}
	=
	2r(r+2b)\left[a^2+r^2-2(1+b)r\right],
\end{equation}

\vspace{0.2cm}
and therefore

\begin{equation}
	\left.
	\frac{E_{\rm CM}^2}{2M_0^2}
	\right|_{L_1=L_2=0}
	=
	\frac{
		2r(r+2b)\left[a^2+r^2-2(1+b)r\right]
	}{
		r(r+2b)\left[a^2+r^2-2(1+b)r\right]
	}
	=
	2 .
\end{equation}

\vspace{0.2cm}
Thus,

\begin{equation}
	\left.
	E_{\rm CM}
	\right|_{L_1=L_2=0}
	=
	2M_0 ,
\end{equation}

\vspace{0.2cm}
\noindent
which is finite and agrees with the finite radial limit obtained by Hussain. The difference between the two results is therefore not due to an error in Hussain's calculation, but to the presence or absence of the critical angular-momentum channel. The relevant near-horizon quantity is

\begin{equation}
	X_i=E_i-\Omega_H L_i .
\end{equation}

\vspace{0.2cm}
\noindent
The BSW critical condition 

\begin{equation}
	X_i=0
	\quad\Longleftrightarrow\quad
	L_i=L_c=\frac{E_i}{\Omega_H}.
\end{equation}

\vspace{0.2cm}
For the extremal EMDA black hole,

\begin{equation}
	a_E=r_E=1+b,
	\qquad
	\Omega_E=\frac{1}{2(1+2b)},
	\qquad
	L_c=2(1+2b)E_i .
\end{equation}

\vspace{0.2cm}
\noindent
In Hussain's radial sector, $L_i=0$, and hence $X_i=E_i>0$. Thus neither particle is critical, the near-horizon cancellation is complete,
and $E_{\rm CM}$ remains finite. By contrast, the divergence found in the
present work arises only when the angular-momentum degree of freedom is retained
and one particle is tuned to $L_i=L_c$. Therefore Hussain's result is not
contradicted; it is recovered in the radial/noncritical sector, while the BSW
divergence belongs to the distinct critical angular-momentum sector.


\end{document}